\documentstyle[11pt]{article}

\newlength{\bredde}
\def\slash#1{\settowidth{\bredde}{$#1$}\ifmmode\,\raisebox{.15ex}{/}
\hspace*{-\bredde} #1\else$\,\raisebox{.15ex}{/}\hspace*{-\bredde} #1$\fi}
\newcommand{\beq}{\begin{equation}}
\newcommand{\eeq}{\end{equation}}
\newcommand{\ba}{\begin{array}{ccc}}
\newcommand{\ea}{\end{array}}
\newcommand{\nn}{\nonumber}
\newcommand{\noi}{\vspace{12pt}\noindent}
\newcommand{\lG}{\raise.3ex\hbox{$\stackrel{\leftarrow}{G}$}}
\newcommand{\lU}{\raise.3ex\hbox{$\stackrel{\leftarrow}{U}$}}
\newcommand{\lP}{\raise.3ex\hbox{$\stackrel{\leftarrow}{{\cal P}}$}}
\newcommand{\leta}{\raise.3ex\hbox{$\stackrel{\leftarrow}{\eta}$}}
\newcommand{\lOmega}{\raise.3ex\hbox{$\stackrel{\leftarrow}{\Omega}$}}
\newcommand{\ldr}{\raise.3ex\hbox{$\stackrel{\leftarrow}{\delta^r}$}}

\def\beqn{\begin{eqnarray}}
\def\eeqn{\end{eqnarray}}

\def\gtwid{\raise.3ex\hbox{$>$\kern-.75em\lower1ex\hbox{$\sim$}}}
\def\ltwid{\raise.3ex\hbox{$<$\kern-.75em\lower1ex\hbox{$\sim$}}}
\def\l{\label}
\def\r{\ref}
\def\la{\lambda}
\def\ga{\gamma}
\def\om{\omega}
\def\tb{{\mbox{\bf t}}}
\begin{document}
\topmargin -1.4cm
\oddsidemargin -0.8cm
\evensidemargin -0.8cm
\title{\Large{{\bf On Finite-Volume Gauge Theory Partition Functions}}}

\vspace{1.5cm}

\author{~\\{\sc G. Akemann}\\Max-Planck-Institut f\"ur Kernphysik\\
Saupfercheckweg 1\\D-69117 Heidelberg\\Germany\\~\\and\\~\\
{\sc P.H. Damgaard}\\
The Niels Bohr Institute\\ Blegdamsvej 17\\ DK-2100 Copenhagen {\O}\\
Denmark}
\date{} 
\maketitle
\vfill
\begin{abstract}
We prove a Mahoux-Mehta--type theorem for finite-volume
partition functions of $SU(N_c\geq 3)$ gauge theories coupled to fermions
in the fundamental representation. The large-volume limit is
taken with the constraint $V << 1/m_{\pi}^4$. The theorem allows one to
express any k-point correlation function of the microscopic Dirac 
operator spectrum entirely in terms of the 2-point function.
The sum over topological charges of the gauge fields can be 
explicitly performed for these k-point correlation functions.
A connection to an integrable KP hierarchy, for which the 
finite-volume partition function is a $\tau$-function, is 
pointed out. Relations between the effective partition functions
for these theories in 3 and 4 dimensions are derived.
We also compute analytically, and entirely from finite-volume
partition functions, the microscopic spectral density of the 
Dirac operator in $SU(N_c)$ gauge theories coupled to quenched
fermions in the adjoint representation. The result coincides
exactly with earlier results based on Random Matrix Theory.  
\end{abstract}
\vfill

\begin{flushleft}
NBI-HE-99-41 \\
hep-th/9910190
\end{flushleft}
\thispagestyle{empty}
\newpage

\section{Introduction}

\noi
While it has been known for some years that Random Matrix Theory provides
universality classes that describe the microscopic spectrum of the Dirac 
operator in theories with spontaneous breaking of chiral symmetry 
\cite{SV,V,ADMN}, it is only more recently that the precise relationship
is become unraveled \cite{AD,OTV,TV}. The essential point is that the
Random Matrix Theory partition function, in a particular scaling limit,
becomes {\em identical} to the effective field theory partition function
in an analogous scaling limit \cite{SV,HV}. 
This scaling limit is best thought of in finite-size scaling terms, as
it requires sending the space-time volume $V$ to infinity with the
constraint that $V \ll 1/m_{\pi}^4$, where $m_{\pi}$ generically indicates
the pseudo-Goldstone masses. At the level of fermion masses $m_i$, one keeps
the product $\mu_i= m_i\Sigma V$ fixed, $\Sigma$ being the infinite-volume
chiral condensate. Using the universality proof of ref. \cite{ADMN} one 
easily establishes that the relation between the partition functions holds
universally, independently of the chosen Random Matrix Theory potential
\cite{Dtalk}. The established identity between the two partition functions
is not sufficient to establish that the microscopic Dirac operator spectrum 
can be computed in Random Matrix Theory, but a series of surprisingly
simple relations that express all microscopic spectral correlators of
on Random Matrix Theory in terms of the universal partition functions
\cite{AD} indicate that one must be able to compute the microscopic
Dirac operator spectrum directly from an effective field theory that
is suitably extended by additional fermionic species. The supersymmetric
formulation of partially quenched effective lagrangians provides an
analytical framework where this can be established \cite{OTV,TV}. This
is an important point, because it {\em proves}, starting directly from
the effective field theory partition function that the microscopic Dirac
operator spectrum coincides with that obtained universally from Random Matrix
Theory. 

\noi
In the derivation of microscopic spectral correlators of the Dirac
operator one makes efficient use of the fact that one can insert factors
of unity inside the field theory path integral by means of cancelling
fermionic and bosonic degrees of freedom. Thus, for the one-point spectral
function, the spectral density itself, it suffices to insert one such
pair, and hence consider a partially quenched field theory partition function 
of the form
\beq
Z_{\nu} ~=~ \left(\prod_{f=1}^{N_{f}} m_f^{\nu}\right)\left(
\frac{m^F_{v}}{m^B_{v}}\right)^{\nu}
\int\! [dA]_{\nu}
~\frac{\det(i\slash{D} - m^F_{v})}{\det(i\slash{D} - 
m^B_{v})}\prod_{f=1}^{N_{f}}
\det(i\slash{D} - m_f) ~e^{-S_{YM}[A]} ~.
\eeq
Here $\nu$ is the topological charge, and the additional (``quenched'')
fermion-boson species have masses $m^F_{v}, m^B_{v}$ that eventually are
taken to be equal (which makes the two determinants cancel). It is also
clear \cite{OTV} that in order to derive higher $k$-point spectral correlation
functions one will have to insert $k$ such additional factors, which pairwise
cancel in the end:
\beq
Z_{\nu} ~=~ \left(\prod_{f=1}^{N_{f}} m_f^{\nu}\right)\left(
\prod_{i=1}^k \frac{m^F_{vi}}{m^B_{vi}}\right)^{\nu}
\int\! [dA]_{\nu}~\prod_{j=1}^k
\frac{\det(i\slash{D} - m^F_{vj})}{\det(i\slash{D} - m^B_{vj})}
\prod_{f=1}^{N_{f}}\det(i\slash{D} - m_f) ~e^{-S_{YM}[A]} ~.
\eeq

\noi
Let us for clarity here briefly restrict ourselves to the universality
class of the chiral Unitary Ensemble (chUE) in Random Matrix Theory language,
which is the appropriate universality class for SU($N_c\geq 3$) gauge
groups with $N_f$ fermions in the fundamental representation (as was
implicitly assumed when we wrote down the partially quenched partition
functions above).
To obtain the spectral correlation function from the effective 
finite-volume partition
function one needs to take a discontinuity at a cut in a $k$th order
chiral susceptibility. Using the technique of the first of ref. \cite{OTV}
it should be possible to rewrite this, at the level of the effective
lagrangian, in terms of a partition function extended with $2k$ additional
species, all of purely imaginary masses. While this has not yet been
explicitly established beyond the one-point function, it seems beyond any
doubt that it will be possible to carry such a program through. The reason
one can say this with such confidence is that it is known, if one
accepts the use of Random Matrix Theory for {\em all} spectral 
correlators, that the $k$-point spectral function in the chUE universality
class can be be written \cite{AD}
\beq
\rho_S^{(\nu)}(\xi_1,\ldots,\xi_k;\{\mu\}) 
= C_2^{(k)} 
\prod_{i=1}^k\left( |\xi_i|\
\prod_{f=1}^{N_f}(\xi_i^2+\mu_f^2)\right)
\prod_{j<l}^k(\xi_j^2-\xi_l^2)^2 
\frac{{\cal Z}_{\nu}^{(N_{f}+2k)}
(\{\mu\},\{i\xi_1\},\ldots, \{i\xi_k\})}
{{\cal Z}_{\nu}^{(N_{f})}(\{\mu\})} ,
\label{corrft}
\eeq  
where on the right hand side
each additional mass $i\xi_j$ is two-fold degenerate.
Indeed, the supersymmetric coset needed to derive the $k$-point spectral
function is $Gl(N_f+k|k)$ \cite{OTV}, which precisely involves a
finite-volume partition function of $k+k$ additional species.

\noi
However, it is known from Random Matrix Theory that there exists a different,
and far more compact, expression for the spectral $k$-point function:
\beq
\rho_S^{(\nu)}(\xi_1,\ldots,\xi_k;\{\mu\}) ~=~
\det_{a,b} K_S^{(\nu)}(\xi_a,\xi_b;\{\mu\}) ~,
\label{correlchUE}
\eeq 
where the microscopic kernel $K_S^{(\nu)}(\xi_a,\xi_b;\{\mu_i\})$ also
can be expressed in terms of finite-volume partition functions alone \cite{AD}:
\beq
K_S^{(\nu)}(\xi,\xi';\mu_1,\ldots,\mu_{N_{f}}) ~=~ 
(-1)^{\nu+[N_f/2]}
\sqrt{\xi\xi'}\prod_{f=1}^{N_{f}}
\sqrt{(\xi^2+\mu_f^2)(\xi'^2+\mu_f^2)}~\frac{
{\cal Z}_{\nu}^{(N_{f}+2)}(\{\mu\},i\xi,i\xi')}{
{\cal Z}_{\nu}^{(N_{f})}(\{\mu\})} ~.\label{mf}
\eeq
The formula (\ref{correlchUE}) is just one out of three compact expressions
for the $k$-point spectral correlators of all three classical matrix
ensembles, which we generically (although their history date much further
back) shall denote Mahoux-Mehta relations \cite{MM}. Taken together with
the expressions (\ref{corrft}) and (\ref{mf}) it implies a surprising
identity for the partition function, which was noted in the last of
reference \cite{AD}, and which we shall denote Consistency Condition I
(to be stated in precise form in section 2 below).

\noi
Although Consistency Condition I is valid without any doubt, it has been 
derived through the rather tortuous route of going through a Random Matrix 
Theory representation for the partition function. It is of interest to
see if this identity can be proven {\em directly} from the effective partition
function itself, without recourse to Random Matrix Theory. In this way
one can logically replace the expression for the $k$-point spectral 
correlation function derived through the supersymmetric technique
(\ref{corrft}) by the much more compact Master Formula (\ref{mf}) and
the relation (\ref{correlchUE}), without having to make use the
Random Matrix Theory formulation. One of the purposes of this paper
is to provide such a direct algebraic proof.

\noi
Random Matrix Theory implies a number of other identities among the
effective finite-volume partition functions, that cannot easily be guessed 
from these partition functions themselves. One of these, which we shall
denote Consistency Condition II below, can be derived in the Random
Matrix Theory language from the relationship between orthogonal polynomials
and the kernel. We shall prove this relation, too. Curiously, if read
conversely this one single relation allows one to derive unambiguously 
the effective partition function for any number of flavors $N_f$, starting with
just two ``boundary conditions'', such as the effective partition function
for $N_f=0$ and $N_f=1$ (which are both trivial).

\noi
Yet other partition function identities arise from the rather simple
relationship between orthogonal polynomials in the chUE and UE Random Matrix
Theories, relevant for QCD-like theories in 4 and 3 space-time dimensions,
respectively. As we shall show in this paper (see section 4), these
relations imply surprising identities among the well-known Harish-Chandra
integral and the external field problem, both for gauge groups 
$U(N_f)$.

\noi
The existence of a long list of ``miraculous'' identities involving
the effective partition function that is relevant for the microscopic Dirac 
operator spectrum is undoubtedly related to the fact that this partition
function can be written as a $\tau$-function of the integrable
KP hierarchy, as we shall discuss in section 5. The connection between
the microscopic Dirac operator spectrum and this integrable system is
certainly interesting in its own right, and may in addition be used to shed new
light on the universal analytical expressions that have been obtained.
It is also intriguing that this connection suggests that the effective
theories relevant for describing the microscopic part of the Dirac operator 
spectrum may be ``topological'' in the sense of Witten \cite{Witten},
$i.e.$ having an entirely different formulation in terms of a BRST-exact
field theory action.
Intuitively this is perhaps understandable from the fact that the
effective partition function in this regime stems from the zero modes of
the pseudo-Goldstone bosons alone, with no kinetic energy contribution.
No degrees of freedom can therefore propagate in this effective theory.

\noi
Having established what we call Consistency Condition I, we get as a
by-product the $\nu$-dependent normalization factor in front of the
spectral $k$-point function. With this factor explicitly known, one can
then perform the sum over topological charges $\nu$ of the gauge field
configurations to provide a compact expression for this $k$-point
function in terms of the full effective partition functions that are already 
summed over topological charges. We do this simple derivation of the
full $k$-point function in section 6. In section 7 we turn to one of the 
other universality classes, labeled chSE, which is the one relevant for 
SU($N_c\geq 2$) gauge theories 
with $N_f$ fermions in the adjoint representation. The analogous formula
for the microscopic spectral density, as derived through Random Matrix
Theory \cite{AD}, involves the partition function of four additional
species. We check that the resulting formula is correct at least in the
fully quenched case by explicitly evaluating the relevant effective
partition function for four flavors. Finally, section 8 contains our
brief conclusions. Some technical details are relegated to the appendices. 

\noi
Before starting with the main part of the paper let us fix the notation for
clarity. The finite-volume partition function for QCD-like theories in even
space-time dimensions can be written \cite{JSV}
\beq
{\cal Z}_\nu^{(N_f)} (\{\mu\}) ~=~ \det A(\{\mu\})/\Delta(\{\mu^2\})
\label{ZUE}
\eeq
where the matrix $A$ is defined either by
\beq
A(\{\mu\})_{ij}=\mu_i^{j-1}I_{\nu+j-1}(\mu_i)~~,~~~~i,j=1,\ldots,N_f ~,
\label{Adef}
\eeq
or alternatively (by making use of standard Bessel function identities,
and invariance properties of the determinant),
\beq
A(\{\mu\})_{ij}=\mu_i^{j-1}I_{\nu}^{(j-1)}(\mu_i)~~,~~~~i,j=1,\ldots,N_f ~.
\label{alt}
\eeq
The denominator is given by the Vandermonde determinant of the squared masses
\beq
\Delta (\{\mu^2\})\ \equiv\ \prod_{i>j}^{N_f}(\mu_i^2-\mu_j^2)
\ =\ \det_{i,j}\left[ (\mu_i^2)^{j-1}\right] \ . 
\label{Vandermonde}
\eeq
In the literature a different sign is very often chosen inside the product,
which leads to an overall $N_f$-dependent sign compared with the determinant
$\det_{i,j}[ (\mu_i^2)^{j-1}] 
= (-1)^{N_f(N_f-1)/2}\prod_{i<j}^{N_f}(\mu_i^2-\mu_j^2)$.
This just gives an overall sign in the partition function eq. (\r{ZUE}), which
normally is irrelevant. Let us stress, however, that only with the 
definition above the partition function ${\cal Z}_\nu^{(N_f)} (\{\mu\})$
is a positive quantity. We choose the form (\ref{Vandermonde}) in what
follows.

\setcounter{equation}{0}
\section{Consistency Condition I}

\noi
First of all let us recall the consistency condition relating 
the determinant of partition functions with $N_f+2$ flavors 
${\cal Z}_\nu^{(N_f+2)}$ and a partition functions with $N_f+2k$ flavors 
as it has been stated in \cite{AD}. Using properties of determinants
eq. (11) of the third ref. \cite{AD} is equivalent to
\beq
\det_{1\leq a,b\leq k}\left[
\frac{{\cal Z}_{\nu}^{(N_{f}+2)}(\{\mu\},\xi_a,\xi_b)}
{{\cal Z}_{\nu}^{(N_{f})}(\{\mu\})}\right]
\ =\  
\prod_{i<j}^k\left(\xi_i^2-\xi_j^2\right)^2 ~
\frac{{\cal Z}_{\nu}^{(N_{f}+2k)}
(\{\mu\},\xi_1,\xi_1,\ldots,\xi_k,\xi_k)}
{{\cal Z}_{\nu}^{(N_{f})}(\{\mu\})} \ .
\l{cond1old}
\eeq
The proportionality constant that remained undetermined
in \cite{AD} is therefore fixed\footnote{There is a misprint in
eq. (11) of the third paper in ref. \cite{AD}, where 
the overall factor should read $C_2^{(k)}(-1)^{k(\nu+[N_f/2])}$.}
as we will see below to be
$C_2^{(k)}=(-1)^{k(\nu+[N_f/2])}$.

\noi
In order to prove eq. (\r{cond1old}) we make a more general statement
which will be more easy to prove  due to the lack of degeneracy of the
fermion masses, which is present on the right hand side of eq. (\r{cond1old}):
\beq
\det_{1\leq a,b\leq k}\left[
\frac{{\cal Z}_{\nu}^{(N_{f}+2)}(\{\mu\},\xi_a,\eta_b)}
{{\cal Z}_{\nu}^{(N_{f})}(\{\mu\})}\right]
\ =\  
\prod_{i<j}^k(\xi_i^2-\xi_j^2)(\eta_i^2-\eta_j^2) ~
\frac{{\cal Z}_{\nu}^{(N_{f}+2k)}
(\{\mu\},\xi_1,\ldots,\xi_k,\eta_1,\ldots,\eta_k)}
{{\cal Z}_{\nu}^{(N_{f})}(\{\mu\})} \ .
\l{cond1}
\eeq
Taking the degeneracy limit $\eta_i=\xi_i$, $i=1,\ldots,k$ , we recover the 
original claim eq. (\r{cond1old}).

\noi
We will not be able to give a proof of the above statement in full generality.
In a first step we will prove that it holds in the asymptotic regime
where $\mu_f, \xi_i, \eta_j\to\infty$. In particular this fixes the 
proportionality constant in eq. (\r{cond1}) 
which may depend on $N_f$ and $\nu$. In a second step we can prove
eq. (\r{cond1}) for finite arguments and any $k\in$ N in the quenched case
$N_f=0$ in an arbitrary topological sector $\nu$. Using the established
flavor-topology duality \cite{Jac} in this finite-volume scaling
regime, we have then automatically also proven the identity for {\em any}
number of massless flavors $N_f$ in a sector of {\em any} topological
charge $\nu$. The statement eq. (\r{cond1}) is this case reads

\noi
{\sc Theorem} - Consistency Condition I (massless):

\noi
Let ${\cal Z}_\nu^{(N_f)} (\{\mu\})= \det A(\{\mu\})/\Delta(\{\mu^2\})$
be defined as in eqs.(\r{ZUE})--(\r{Vandermonde}). For $N_f$ massless flavors 
the following identity holds, where we have chosen the constant
${\cal Z}_{\nu}^{(N_{f})}(0)$ to be unity:
\beq
\det_{1\leq a,b\leq k}\left[
{\cal Z}_{\nu}^{(N_{f}+2)}(\xi_a,\eta_b)\right]
\ =\  
\prod_{i<j}^k(\xi_i^2-\xi_j^2)(\eta_i^2-\eta_j^2) ~
{\cal Z}_{\nu}^{(N_{f}+2k)}
(\xi_1,\ldots,\xi_k,\eta_1,\ldots,\eta_k)\ .
\l{cond1m}
\eeq

\noi
{\sc Proof}: (i) asymptotic region (massive, eq. (\r{cond1})):

\noi
Inserting the explicit form of the partition function into eq. (\r{cond1}) 
and factorizing out common parts of the Vandermonde determinants
eq. (\r{cond1}) can be brought into the following form:
\beq
\det_{1\leq a,b\leq k}\left[\frac{1}{\eta_b^2-\xi_a^2}
\frac{\det A(\{\mu\},\xi_a,\eta_b)}
{\det A(\{\mu\})}\right]
\ =\  
\prod_{a,b=1}^k\frac{1}{(\eta_a^2-\xi_b^2)}\ \ 
\frac{\det A(\{\mu\},\xi_1,\ldots,\xi_k,\eta_1,\ldots,\eta_k)}
{\det A(\{\mu\})} \ .
\l{cond1A}
\eeq
We can now apply the asymptotics of Bessel functions
\beq
\lim_{x\to\infty} I_n(x)\ =\ \frac{e^x}{\sqrt{2\pi x}}(1+O(x^{-1})) \ ,
\eeq
which leads to 
\beq
\det A(\{\mu\},\{\xi\},\{\eta\}) \ \longrightarrow\ 
\prod_{i=1}^k\frac{e^{\xi_i+\eta_i}}{\sqrt{2\pi \xi_i\eta_i}}
\prod_{f=1}^{N_f}\frac{e^{\mu_f}}{\sqrt{2\pi \mu_f}}\ 
\Delta(\{\mu\},\{\xi\},\{\eta\})
\l{limA}
\eeq
where we have taken all arguments to infinity and where
$\Delta(\{\mu\},\{\xi\},\{\eta\})$ is now the Vandermonde of
the {\it unsquared } sets of variables. Inserting the result
eq. (\r{limA}) into eq. (\r{cond1A}) we obtain for the left hand side (l.h.s.):
\beqn
\mbox{l.h.s.}&\longrightarrow& 
\det_{1\leq a,b\leq k}\left[\frac{1}{\eta_b^2-\xi_a^2}
\frac{e^{\xi_a+\eta_b}}{\sqrt{2\pi \xi_a\eta_b}}
\frac{\Delta(\{\mu\},\xi_a,\eta_b)}{\Delta(\{\mu\})}\right]\nn\\
&=& \prod_{i=1}^k\frac{e^{\xi_i+\eta_i}}{\sqrt{2\pi \xi_i\eta_i}}
\prod_{f=1}^{N_f}\prod_{a=1}^{k}(\mu_f-\xi_a)(\mu_f-\eta_a)
\det_{1\leq a,b\leq k}\left[\frac{1}{\eta_b+\xi_a}\right]
\eeqn
and for the right hand side (r.h.s.)
\beqn
\mbox{r.h.s.}&\longrightarrow&\prod_{a,b=1}^k\frac{1}{(\eta_a^2-\xi_b^2)}
 \prod_{i=1}^k\frac{e^{\xi_i+\eta_i}}{\sqrt{2\pi \xi_i\eta_i}}
\frac{\Delta(\{\mu\},\{\xi\},\{\eta\})}{\Delta(\{\mu\})} \nn\\
&=&\prod_{i=1}^k\frac{e^{\xi_i+\eta_i}}{\sqrt{2\pi \xi_i\eta_i}}
\prod_{f=1}^{N_f}\prod_{a=1}^{k}(\mu_f-\xi_a)(\mu_f-\eta_a)
\frac{\prod_{a<b}^{k}(\xi_a-\xi_b)(\eta_a-\eta_b)}
{\prod_{a,b=1}^k(\eta_a+\xi_b)} \ .
\eeqn
Putting both sides together and dropping common factors
we obtain
\beq
\det_{1\leq a,b\leq k}\left[\frac{1}{\eta_b+\xi_a}\right] \ =\ 
\frac{\prod_{a<b}^{k}(\xi_a-\xi_b)(\eta_a-\eta_b)}
{\prod_{a,b=1}^k(\eta_a+\xi_b)} \ ,
\l{Cauchy}
\eeq
which is nothing else than Cauchy's Lemma.
The asymptotic analysis performed so far determines the mass independent
overall proportionality constant in eq. (\r{cond1}) to be unity.

\noi
(ii) proof for $N_f$ massless fermions with arbitrary $\nu$ 
(eq. (\r{cond1m})): 

\noi
Due to the flavor-topology duality it is sufficient to prove the statement
for $N_f=0$ with arbitrary $\nu\neq 0$, and then shifting $\nu\to N_f+\nu$.
Let us first give an outline of the proof. We will proceed with the simplified
version eq. (\r{cond1A}) of the theorem for $N_f=0$,
where the Vandermonde determinants have been already cancelled. 
In a first step we shall further simplify
$\det A(\{\xi\},\{\eta\})$ 
using Lemma 1 in the Appendix \r{A}. There it is shown that 
$\det A(\{\xi\},\{\eta\})$ is 
given by a determinant similar to a Vandermonde containing
powers of $\xi_i$ and $\eta_i$ as well as first derivatives with respect to 
these variables acting on a product of Bessel functions of 
the same index $\nu$. In this form we can prove the theorem
by performing a Laplace expansion of the right hand side of eq. (\r{cond1A}) 
and using Lemma 2 for the left hand side.

\noi
Starting with the right hand side the determinant can be rewritten using 
Lemma 1:
\beq
\det A(\xi_1,\ldots,\xi_k,\eta_1,\ldots,\eta_k)  \ =\  
(-1)^{\frac{k(k-1)}{2}}
\left| \begin{array}{cc} B(\xi)& C(\xi)\\ B(\eta)& C(\eta) \end{array} \right|
\ \prod_{i=1}^k I_{\nu}(\xi_i)I_{\nu}(\eta_i)
\eeq
where $B(\xi)$ is the matrix of the Vandermonde determinant of squared 
arguments $B(\xi)_{ij}=(\xi_i^2)^{j-1}$ and $C(\xi)$ is the same as $B$ with an
additional $\xi_i\vec{\partial}_{\xi_i}$ in each row, which is defined to
act only on the Bessel functions. If we perform a Laplace 
expansion with respect to the first $k$ columns into $k\times k$ blocks we can 
use the fact that the resulting determinants $\det B\det C$ can all be 
rewritten as the product of two Vandermonde determinants times $k$ linear
differential operators that can be taken out of $\det C$. We obtain
\beqn
\mbox{r.h.s.}&=& (-1)^{\frac{k(k-1)}{2}}
\prod_{a,b=1}^k\frac{1}{(\eta_a^2-\xi_b^2)}\ 
\sum_\sigma (-1)^\sigma \Delta(x^2_{\sigma(1)},\ldots,x^2_{\sigma(k)})
\ \times\nn\\
&&\times\ \Delta(x^2_{\sigma(k+1)},\ldots,x^2_{\sigma(2k)})
\prod_{i=k+1}^{2k} x_{\sigma(i)}\partial_{x_{\sigma(i)}}
 \prod_{j=1}^{2k} I_{\nu}(x_j) 
\l{rhsfinal}
\eeqn
where we have renamed 
\beq
x_i\ =\ \xi_i\ , \ \  x_{i+k}=\eta_i\ \  \mbox{for} \ \ i=1,..,k\ .
\l{xieta}
\eeq
The permutations $\sigma$ run over all possible 
$\scriptsize{\left(\!\begin{array}{c}2k\\k\end{array}\!\right)}$ 
permutations to put $2k$
variables into 2 sets of $k$ variables with ordered indices
respecting $\sigma(1)<\ldots<\sigma(k)$ and 
$\sigma(k+1)<\ldots<\sigma(2k)$. The sign of the permutation is
defined by $(-1)^\sigma=(-1)^{1+\ldots+k+\sigma(1)+\ldots+\sigma(k)}$ .

\noi
The building blocks of the left hand side of eq. (\r{cond1A}) are the 
$2\times 2$ determinants 
\beq
\det A(\xi_a,\eta_b)\ =\ (\eta_b\partial_{\eta_b}-\xi_a\partial_{\xi_a})
 I_\nu(\xi_a)I_\nu(\eta_b)
\eeq
where we have used again Lemma 1 eq. (\r{la1}). 
Therefore we can rewrite the left hand side as
\beq
\mbox{l.h.s.}\ =\ \det_{1\leq a,b\leq k}\left[\frac{1}{\eta_b^2-\xi_a^2}\det 
A(\xi_a,\eta_b)\right]
\ =\ \det_{1\leq a,b\leq k}\left[\frac{1}{\eta_b^2-\xi_a^2}
\left(\eta_b\vec{\partial}_{\eta_b}-\xi_a\vec{\partial}_{\xi_a}\right)\right]
\prod_{i=1}^k I_\nu(\xi_i)I_\nu(\eta_i) \ ,
\l{lhsI}
\eeq
with the derivatives only acting on the Bessel functions. 

\noi
Using again the fact that determinants differing only by a single column
can be added, eq. (\r{lhsI}) can be rewritten as a sum of determinants
containing only one derivative $\eta_b\vec{\partial}_{\eta_b}$ or
$\xi_a\vec{\partial}_{\xi_a}$, $a=1,\ldots,k$. Due to the structure of the
determinant in the columns with  $\eta_b\vec{\partial}_{\eta_b}$ the 
derivative can be taken out as a common factor. After reordering 
columns we end up with
\beq
\det_{1\leq a,b\leq k}\left[\frac{1}{\eta_b^2-\xi_a^2}
\left(\eta_b\vec{\partial}_{\eta_b}-\xi_a\vec{\partial}_{\xi_a}\right)\right]
\ =\ \sum_{i=0}^k \sum_{\ga_i} (-1)^{\ga_i+i} \det G(i;\ga_i)
\prod_{j=i+1}^k \eta_{\ga_i(j)}\partial_{\eta_{\ga_i(j)}} \ ,
\eeq
where the matrices $G(i;\ga_i)$ are defined in Lemma 2 eq. (\r{Gidef}) 
to contain derivatives
$\xi_a\vec{\partial}_{\xi_a}$ in the first $i$ columns. The $\ga_i$ are all
$\scriptsize{\left(\!\begin{array}{c}k\\i\end{array}\!\right)}$ 
permutations for different $G(i;\ga_i)$ with $i$ fixed, where the ordering is 
such that $\ga_i(1)<\ldots<\ga_i(i)$  and $\ga_i(i+1)<\ldots<\ga_i(k)$. 
The factor $(-1)^i$ stems from taking out the common factor minus one of all
columns containing derivatives with respect to the $\xi$'s.

\noi
We can now apply Lemma 2 where $\det G(i;\ga_i)$ is evaluated by Laplace
expansion and Cauchy's Lemma, to obtain
\beqn
&&\det_{1\leq a,b\leq k}\left[\frac{1}{\eta_b^2-\xi_a^2}
\left(\eta_b\vec{\partial}_{\eta_b}-\xi_a\vec{\partial}_{\xi_a}\right)\right]
\ =\nn\\
&&=(-1)^{\frac{k(k-1)}{2}}
\prod_{a,b=1}^k\frac{1}{(\eta_a^2-\xi_b^2)}\
\sum_{i=0}^k \sum_{\ga_i, \bar{\ga}_i} (-1)^{\ga_i+\bar{\ga}_i+i}
\Delta(\xi^2_{\bar{\ga}_i(1)},\ldots,\xi^2_{\bar{\ga}_i(k-i)},
\eta^2_{\ga_i(1)},\ldots,\eta^2_{\ga_i(i)}) \times\nn\\
&&\ \ \times\ \Delta(\xi^2_{\bar{\ga}_i(k-i+1)},\ldots,\xi^2_{\bar{\ga}_i(k)},
\eta^2_{\ga_i(i+1)},\ldots,\eta^2_{\ga_i(k)}) 
\prod_{j=k-i+1}^k \xi_{\bar{\ga}_i(j)}\partial_{\xi_{\bar{\ga}_i(j)}}
\prod_{l=i+1}^k \eta_{\ga_i(l)}\partial_{\eta_{\ga_i(l)}}
\l{lhsfinal}
\eeqn
This equation is nothing else than the operator in eq. (\r{rhsfinal}) acting on
$\prod_{i=1}^k I_\nu(\xi_i)I_\nu(\eta_i)$, where here
in the first sum the number of $\xi$'s in the first Vandermonde 
is made explicit to be $i$.

\noi
To see this we observe that in eq. (\r{lhsfinal}) we have 
$
\sum_{i=0}^k
\scriptsize{\left(\!\begin{array}{c}k\\i\end{array}\!\right)^2}=
\scriptsize{\left(\!\begin{array}{c}2k\\k\end{array}\!\right)}
$
different terms, which matches to the number of permutations
in eq. (\r{rhsfinal}). In order to map individual permutations including signs
we use again the notation eq. (\r{xieta}). The index $\ga(j)$ of the $\eta$'s
thus changes according to
\beq
\ga(j)\ \rightarrow\ k+\ga(j)\ \equiv\ \sigma(j+k)\ ,
\eeq
which implies$
(-1)^{\ga(1)+\ldots+\ga(i)} \ =\ (-1)^{\sigma(1+k)+\ldots+\sigma(i+k) -ik}$.
Consequently we obtain
\beqn
(-1)^{\ga_i\ +\ \bar{\ga}_i\ +\ i} 
&=& (-1)^{\frac{(i)(i+1)}{2}
\ +\ \sum_{l=1}^{k-i}\sigma_i(k-l)\ -\ ik
\ +\ \frac{(k-i)(k-i+1)}{2}
\ +\sum_{l=1}^{k-i}\bar{\ga}_i(l)
\ +\ i}  \nn\\
&=& (-1)^{\frac{k(k+1)}{2} \ +\ \sum_{j=1}^{k}\sigma_i(j)} \ =\ (-1)^{\sigma_i}
\ ,
\eeqn
where $\sigma_i$ is a permutation with $i$ $\xi$'s with ordered indices
$\{ \bar{\ga}_i(j)=\sigma_i(j);\ j\!=\!1,\ldots,k-i\}\in\{ 1,\ldots,k\}$
and $(k-i)$ $\eta$'s with ordered indices 
$\{ \sigma_i(k+j);\ j\!=\!1,\ldots,i\}\in\{ k+1,\ldots,2k\}$
as in the first Vandermonde in eq. (\r{rhsfinal}).
We have thus completed the matching of both sides of theorem 
eq. (\r{cond1m}).

\setcounter{equation}{0}
\section{Consistency Condition II}

In this section we will prove the following theorem relating partition 
functions with $N_f$, $N_f+1$ and $N_f+2$ massive flavors, as it has
been stated in the last of reference \cite{AD}.

\noi
{\sc Theorem} - Consistency Condition II:

\noi
For ${\cal Z}_\nu^{(N_f)} (\{\mu\})= \det A(\{\mu\})/\Delta(\{\mu^2\})$
as defined in eqs.(\r{ZUE})--(\r{Vandermonde}) it holds
\beqn
{\cal Z}_\nu^{(N_f+2)} (\{\mu\},\xi,\eta) &=& 
\frac{1}{(\xi^2-\eta^2){\cal Z}_\nu^{(N_f)} (\{\mu\})} \times\nn\\
&\times& \left[ \left( (\sum_{i=1}^{N_f}\mu_i\partial_{\mu_i} 
+\xi\partial_{\xi})
{\cal Z}_\nu^{(N_f+1)}(\{\mu\},\xi)\right)
{\cal Z}_\nu^{(N_f+1)}(\{\mu\},\eta) ~-~ 
\left(\xi \leftrightarrow \eta\right)\right]
\l{cond2}
\eeqn

\noi
A few remarks can be made here. Taking the inverse statement, 
the differential equation eq. (\r{cond2}) together with the
boundary conditions ${\cal Z}_\nu^{(0)}=1$ and 
${\cal Z}_\nu^{(1)}(\mu)=I_\nu(\mu)$ can be seen as a generating equation
for all ${\cal Z}_\nu^{(N_f)} (\{\mu\})$. In this way we can actually 
{\em derive} the precise form of the partition function eqs.
(\r{ZUE})--(\r{Vandermonde}) instead of taking it as a starting point.
In ref. \cite{WGW}, an even more compact recursive relation for the
partition functions was derived in the Random Matrix Theory formulation,
using the supersymmetric formalism.

\noi
The consistency condition eq. (\r{cond2}) had been
conjectured in the third reference of \cite{AD} (eq. (16)) 
with an arbitrary constant, which
has been determined here to be $C=1$. This result follows independently
from inserting the asymptotics of the Bessel functions.

\noi
{\sc Proof}:
Before going into the details let us give the general outline of the proof.
We will first investigate the action of the power-counting operator
$\sum_{i=1}^{N_f}\mu_i\partial_{\mu_i}+\xi\partial_{\xi}$
on the partition function ${\cal Z}_\nu^{(N_f+1)}(\{\mu\},\xi)$. Most 
of the outcome will be again proportional to the same partition function
which then drops out due to the antisymmetry with respect to $\xi$ and $\eta$ 
in the bracket in eq. (\r{cond2}). The remainder times the partition functions
${\cal Z}_\nu^{(N_f+1)}(\{\mu\},\eta)$ 
expanded once will then precisely arrange
to the Laplace expansion of the left hand side with respect to the last two 
columns. Throughout the proof we will make use of 3 different Lemmas collected 
in Appendix \r{B}.

\noi
Let us start with the action of the power-counting differential operator
on the Vandermonde in the denominator. It is easy to show that
\beq
(\sum_{i=1}^{N_f}\mu_i\partial_{\mu_i} +\xi\partial_{\xi})\ 
\Delta(\{\mu^2\},\xi^2)\ =\ (N_f+1)N_f\ \Delta(\{\mu^2\},\xi^2)
\eeq
since the operator counts the sum of all powers.
{}From the product rule every factor in $\Delta$ gets differentiated twice
and is thus reproduced with a factor of 2 in front, furthermore there are
$(N_f+1)N_f/2$ such factors. Hence the differentiation of the $\Delta$'s
drops out of eq. (\r{cond2}) due to the $(\xi \leftrightarrow \eta)$
antisymmetry. Inserting the explicit form of the partition function
we obtain
\beq
\det A(\xi,\eta)\ =\ -(\det A)^{-1}
\left[ \left( (\sum_{i=1}^{N_f}\mu_i\partial_{\mu_i} +\xi\partial_{\xi})
\det A(\xi)\right) \det A(\eta)~-~ 
(\xi \leftrightarrow \eta)\right]
\l{cond2det}
\eeq
after cancelling the Vandermonde determinants. Here and in the rest of this 
section we have omitted the dependence of the matrix $A$ on the set 
of variables $\mu_i$. Next we apply the differential operator
to the determinant of $A(\xi)$. 
Since it is linear and the matrix $A$ depends on each 
variable only in one row we can differentiate row-wise, using
\beq
\mu\partial_\mu (\mu^n I_{\nu+n}(\mu)) \ =\ (2n + \nu)\mu^n I_{\nu+n}(\mu)
\ +\ \mu^{n+1} I_{\nu+n+1}(\mu).
\l{dI}
\eeq
In fact this is the only place where the properties of Bessel functions enter,
the rest of the argument being valid for general matrices $A$.
Determinants that only differ by one row can be added and we thus have
\beq
\sum_{i=1}^{N_f}\mu_i\partial_{\mu_i}\det A\ =\ 
\sum_{k=1}^{N_f}\left[2\sum_{L=2}^{N_f}\det A_k^L +\nu\det A
\ +\ \det A_k\right]\ .
\l{DdetA}
\eeq
Here in the last term
$A_k$ is the matrix $A$ with the $k$-th row shifted
to the left by one unit, $(A_k)_{kl}=\mu_k^{l}I_{\nu+l}(\mu_k)$ and $A_k=A$
else, which results from the last term in eq. (\r{dI}). This last term which
can be further simplified using Lemma 3 in Appendix \r{B} will be the only 
term that survives in eq. (\r{cond2det}).
The second term $\nu\det A$ vanishes immediately due to the antisymmetry 
in eq. (\r{cond2det}).

\noi
We will now show that also the first term in eq. (\r{DdetA}) is proportional
to $\det A$ and thus drops out. The matrix $A_k^L$ is defined to be 
the matrix $A$ with the first $L$ entries in the $k$-th row vanishing,
$(A_k^L)_{km}=0$ for $m=1,\ldots,L$ and $A_k^L=A$ else. The sum over $L$
in eq. (\r{DdetA}) thus reproduces the $2n\mu^n I_{\nu+n}(\mu)$ from 
eq. (\r{dI}). Applying Lemma 4 from Appendix \r{B} we obtain
\beq
2\sum_{L=2}^{N_f}\sum_{k=1}^{N_f}\det A_k^L \ =\ 2\sum_{L=2}^{N_f}(N_f-L)\det A
\ =\ (N_f-1)(N_f-2) \det A \ .
\eeq
Together with Lemma 3 Appendix \r{B} we finally obtain
\beq
\sum_{i=1}^{N_f}\mu_i\partial_{\mu_i}\det A\ =\ 
((N_f-1)(N_f-2) +\nu N_f) \det A
+\sum_{j=1}^{N_f}(-1)^{N_f+j}\mu_j^{N_f}
I_{\nu+N_f}(\mu_j) \det A^*_{jN_f}
\l{DdetA2}
\eeq
where $A^*_{jN_f}$ is the matrix $A$ with row $j$ and column $N_f$ missing
(algebraic complement of matrix element $A_{jN_f})$. 
Inserting eq. (\r{DdetA2}) for the derivative of the  
$(N_f+1)\times(N_f+1)$ determinant of $A(\xi)$ in eq. (\r{cond2det})
we obtain 
\beqn
\det A(\xi,\eta) \det A &=& 
\mbox{\huge (} \sum_{j=1}^{N_f}(-1)^{N_f+1+j}\mu_j^{N_f+1}I_{\nu+N_f+1}(\mu_j) 
                             \det A^*_{jN_f+1}(\xi) \nn\\
&&-\ \ \xi^{N_f+1}I_{\nu+N_f+1}(\xi) 
    \det A\mbox{\huge )} \det A(\eta)\ \ 
-\ \ (\xi \leftrightarrow \eta) \ .
\l{cond2exp}
\eeqn
Expanding $\det A(\eta)$ and $\det A(\xi)$ on the right hand side
with respect to the last column and multiplying out
we obtain for the right hand side of eq. (\r{cond2exp})
\beqn
\mbox{r.h.s.}\!\!\!&=&
\!\!\!\!\!\!\!\!\!\sum_{{\scriptsize\ba &i,j=1&\\ &i\neq j&\ea}}^{N_f}
\!\!\!\!\!\!(-1)^{i+j}
        \mu_j^{N_f+1}I_{\nu+N_f+1}(\mu_j) \mu_i^{N_f}I_{\nu+N_f}(\mu_i) 
     \left[\det A^*_{jN_f+1}(\xi) \det A^*_{iN_f+1}(\eta)
         -(\xi \leftrightarrow \eta) \right] \nn\\
+&\mbox{\Huge [}& \!\!\!\!\sum_{j=1}^{N_f}(-1)^{N_f+1+j}
        \mu_j^{N_f+1}I_{\nu+N_f+1}(\mu_j)     
    \left(\eta^{N_f}I_{\nu+N_f}(\eta) \det A^*_{jN_f+1}(\xi)
    - \xi^{N_f}I_{\nu+N_f}(\xi) \det A^*_{jN_f+1}(\eta)\right)\nn\\
&+& \!\!\!\!\sum_{j=1}^{N_f}(-1)^{N_f+1+j}
        \mu_j^{N_f}I_{\nu+N_f}(\mu_j)     \left(
         \xi^{N_f+1}I_{\nu+N_f+1}(\xi) \det A^*_{jN_f}(\eta)
-\eta^{N_f+1}I_{\nu+N_f+1}(\eta)\det A^*_{jN_f}(\xi)\right)\nn\\
&+&\!\!\!\!\left(\xi^{N_f+1}I_{\nu+N_f+1}(\xi) \eta^{N_f}I_{\nu+N_f}(\eta) 
    \ -\  \eta^{N_f+1}I_{\nu+N_f+1}(\eta) \xi^{N_f}I_{\nu+N_f}(\xi)
  \right) \det A \ \mbox{\Huge ]}\det A \nn\\
\l{rhscond2}
\eeqn
We are now ready to apply Lemma 5 from Appendix \r{B} which relates products
of determinants of matrices which just differ by the last two rows. 
The bracket in the first line then reads
\beq
\left[\det A^*_{jN_f+1}(\xi) \det A^*_{iN_f+1}(\eta)
         - (\xi \leftrightarrow \eta) \right]\ =\  
 \det A^*_{iN_f+1;jN_f+2}(\xi,\eta)\det A
\left\{ \ba 
\cdot(+1)& & i<j \\
\cdot(-1)& & i>j \ea \right. 
\eeq
where we have introduced the matrix $A^*_{iN_f+1;jN_f+2}$
where $i$-th and $j$-th row and the last two columns are missing. The obvious 
symmetry $A^*_{iN_f+1;jN_f+2}=A^*_{jN_f+1;iN_f+2}$ leads to a sum over $i<j$
only in the first line of eq. (\r{rhscond2}). Since we can trivially rewrite
$A^*_{jN_f+1}(\xi)=A^*_{iN_f+1;N_f+2N_f+2}(\xi,\eta)$
one can immediately see that eq. (\r{rhscond2}) is nothing else than the 
Laplace expansion of $\det A(\xi,\eta)$ in $2\times2$ times 
$N_f\times N_f$ blocks choosing the last two columns for the expansion.
Thus we have obtained the right hand side of eq. (\r{cond2exp}) and 
completed the proof.

\setcounter{equation}{0}
\section{Relations between QCD$_3$ and QCD$_4$ partition functions}

\noi
Let us first recall how the Consistency Condition I and II of the the previous 
sections have been originally derived in \cite{AD}. 
The important point is that every finite volume partition function
equals a partition function of a massive chiral random matrix model
in the microscopic large-$N$ scaling limit. 
The fact that in matrix models the orthogonal polynomials, the associated 
kernel,
as well as all correlation functions can be expressed in terms of 
matrix model
partition functions translates the well known relations between these
objects to relations among finite volume partition functions.
Consistency Condition II reflects the Christoffel-Darboux identity
between the kernel and the orthogonal polynomials, and Consistency Condition I
is the Mahoux-Mehta relation among $k$-point correlation functions
and the determinant of the kernel.

\noi
In this section we will follow the same reasoning to translate relations
between matrix model quantities for QCD$_3$ and QCD$_4$ to the corresponding 
finite volume partition functions. In contrast to the previous section
we will not provide a proof solely based on finite volume partition 
functions. While the identities are very easily derived in terms of
the Random matrix Theory formulation, they translate, in the microscopic
large-$N$ limit into highly non-trivial relations between group theory
integrals of Harish-Chandra type (for unitary groups) and what can
be called the external field problem for unitary groups. These surprising
identities deserve to be understood in their own right.

\noi
Let us start with a relation between the orthogonal polynomials the 
chUE (QCD$_4$) and those of the UE (QCD$_3$):
\beq
P_{N,\ chUE}^{(N_f,\ \nu=-1/2)} (z^2;\{\mu\}) \ =\ 
P_{2N,\ UE}^{(2N_f)}\ (z;\{\mu\}) \ .
\label{OP}
\eeq
This simple identity is very easily derived from the defintion of the 
two Random Matrix Theories as given in \cite{ADMN2}.
Since in ref. \cite{ADMN2} no explicit use was made of the measure,
eq. (\r{OP}) also holds in the massive case\footnote{In contrast to 
\cite{ADMN2} we have shifted the non-integer part to the left, $\nu=-1/2$, 
in order to deal with a physical (even) number of massive flavors for 
QCD$_3$ on the right hand side.}.
On the right hand side the same $N_f$ masses appear in pairs with opposite 
sign.
Expressing the orthogonal polynomials by partition functions as given in 
the second of ref. \cite{AD} we obtain
\beq
C (i\xi)^{1/2} \frac{{\cal Z}_{\nu\!=\!-1/2}^{(N_f+1)}(\{\mu\},i\xi)}
{{\cal Z}_{\nu\!=\!-1/2}^{(N_f)}(\{\mu\})} \ =\ 
\frac{{\cal Z}_{QCD3}^{(2N_f+1)}(\{\mu\},i\xi)}
{{\cal Z}_{QCD3}^{(2N_f)}(\{\mu\})} \ .
\label{3-4}
\eeq
Here the unknown proportionality constant only reflects the 
choice of normalization for the polynomials, and it can easily be fixed.
Since we have started with an even polynomial the odd-flavor
partition function on the right hand side is given by 
\cite{V}
\beq
{\cal Z}_{QCD3}^{(2N_f+1)}\ =\ \int\! dU \cosh
[{\mbox{\rm Tr}({\cal M}}U\Gamma U^\dagger)]\ ,
\eeq
where ${\cal M}$=diag
$(\mu_1,\ldots,\mu_{N_f},-\mu_1,\ldots,-\mu_{N_f},i\xi)$
and $\Gamma$=diag({\bf 1}$_{N_f}$,-{\bf 1}$_{N_f+1}$). The denominator
is given by the 3-dimensional even-flavor partition function \cite{V},
\beq
{\cal Z}_{QCD3}^{(2N_f)}\ =\ \int\! dU \exp
[{\mbox{\rm Tr}({\cal M}}U\Gamma U^\dagger)]\ ,
\eeq
with  ${\cal M}$=diag
$(\mu_1,\ldots,\mu_{N_f},-\mu_1,\ldots,-\mu_{N_f})$
and $\Gamma$=diag({\bf 1}$_{N_f}$,-{\bf 1}$_{N_f}$). These group integrals
are both of the Harish-Chandra type, while the left hand side of eq.
(\ref{3-4}) is given by an entirely different ratio of unitary group theory 
integrals of the external-field kind. This is the first of such relations.

\noi
The relation between the orthogonal polynomials eq. (\r{OP})
can be exploited furthermore in order to relate also the corresponding kernels
of the chUE and the UE. For convenience, let us introduce the ``wave 
functions'' 
\beq
\Psi_n(\lambda) ~\equiv~ \sqrt{\omega(\lambda)}P_n(\lambda) ~,
\eeq
where $\omega(\lambda)$ is the measure function (so that the wave functions
$\Psi_n(\lambda)$ are orthogonal with respect to a weight of unity).
Now, from eq. (3.11) of ref. \cite{ADMN2} we can use the wave functions
of the UE inside the Christoffel-Darboux identity
for the chUE kernel\footnote{Eq. (\r{OP}) trivially
holds also for the wave functions instead of the polynomials.}.
Doing this, we readily derive the following identities:
\beqn
K_{N,\ chUE}^{(N_f,\ \nu=-1/2)} (z^2,w^2) &=&
\frac{c_{2N}}{z^2-w^2}
\left( \Psi_{2N}^{UE}(z)\ w\ \Psi_{2N-1}^{UE}(w)
\ -\ z\ \Psi_{2N-1}^{UE}(z)\Psi_{2N}^{UE}(w)\right)  \nn\\
&=&\frac{1}{2}\frac{c_{2N}}{(z-w)}
\left( \Psi_{2N}^{UE}(z)\Psi_{2N-1}^{UE}(w)
\ -\ \Psi_{2N-1}^{UE}(z)\Psi_{2N}^{UE}(w)\right)  \nn\\
&&-\ \frac{1}{2}\frac{c_{2N}}{(z+w)}
\left( \Psi_{2N}^{UE}(z)\Psi_{2N-1}^{UE}(w)
\ + \ \Psi_{2N-1}^{UE}(z)\Psi_{2N}^{UE}(w)\right)  \nn\\
&=&\frac{1}{2}\left( K_{2N,\ UE}^{(2N_f)} (z,w) \ +\  
K_{2N,\ UE}^{(2N_f)} (-z,w) \right) \ ,
\label{kernel}
\eeqn
where for clarity we have not explcitly indicated the mass dependence of the 
wave functions. 
The coefficients $c_{2N}$ are defined in ref. \cite{ADMN2}.
We have here made use of the fact that in the
UE the polynomials $P_n(\la)$ are of parity $(-1)^n$.
As one can easily convince oneself the last equation is, despite its
appearance, symmetric in the arguments $z$ and $w$. 
Expressing the kernels in terms of the corresponding 
finite volume partition functions \cite{AD} eq. (\r{kernel}) leads to 
the following relation:
\beq
(-1)^{[\frac{N_f}{2}]-\frac{1}{2}}\sqrt{\xi\om}\ 
\frac{{\cal Z}_{\nu\!=\!-1/2}^{(N_f+2)}(\{\mu\},i\xi,i\om)}
{{\cal Z}_{\nu\!=\!-1/2}^{(N_f)}(\{\mu\})}
\ =\ \frac{1}{4\pi}\left(
\frac{{\cal Z}_{QCD3}^{(2N_f+2)}(\{\mu\},i\xi,i\om)}
{{\cal Z}_{QCD3}^{(2N_f)}(\{\mu\})}
\ - \ \frac{{\cal Z}_{QCD3}^{(2N_f+2)}(\{\mu\},-i\xi,i\om)}
{{\cal Z}_{QCD3}^{(2N_f)}(\{\mu\})} \right)\ .
\eeq
The proportionality constant is known in this case as it can be determined
for both kernels from the matching condition with the microscopic
spectral density.

\noi
Let us finally mention that also in the case of QCD$_3$-like theories
with an {\em odd} number of massive flavors a relation similar to eq.
(\r{kernel})
exists. In ref.\cite{Jesper} the random matrix model kernel for odd-flavored
QDC$_3$ has been derived from a chUE, which in the microscopic scaling limit
reads:
\beq
\frac{\xi+\om}{\sqrt{\xi\om}}
K_{S,\ chUE}^{(N_f,\ \nu=+1/2)}(\xi,\om;\{\mu\}) 
\ =\ 
 K_{S,\ UE}^{(2N_f+1)} (\xi,\om;\{\mu\},0) \ .
\eeq
Inserting again the representation in terms of partition functions \cite{AD},
we obtain the following relation:
\beq
\frac{\xi+\om}{\sqrt{\xi\om}} \ (-1)^{\frac{1}{2}+[\frac{N_f}{2}]}
\frac{{\cal Z}_{\nu\!=\!1/2}^{(N_f+2)}(\{\mu\},i\xi,i\om)}
{{\cal Z}_{\nu\!=\!1/2}^{(N_f)}(\{\mu\})} 
\ =\ 
\frac{1}{2\pi}
\frac{{\cal Z}_{QCD3}^{(2N_f+1+2)}(\{\mu\},0,i\xi,i\om)}
{{\cal Z}_{QCD3}^{(2N_f+1)}(\{\mu\},0)} \ .
\eeq
For more details on the odd-flavor partition function we refer to ref.
\cite{Jesper}.

\noi
We have explicitly checked all of the above relations for a few (small)
number of flavors, starting from the finite-volume partition
functions alone. Apart from those given above, we 
have empirically found other non-linear relations between these 
Harish-Chandra type integrals and those of the unitary external field
problem. These relations, however, do not seem to follow easily
from the Random Matrix Theory formulation.

\setcounter{equation}{0}
\section{The effective QCD$_4$ partition function as a $\tau$-function}

\noi
The identities derived above may have their origin in a surprising
relation to integrable systems, which we shall now briefly discuss.
The starting point is the following observation. Suppose we define
an $N_f\times N_f$ hermitian matrix integral by
\beq
\tau(X) ~=~ \int dY \exp\left[{\mbox{\rm Tr}}[XY + V(Y)]\right] ~.
\label{taumatrix}
\eeq
With $X$ itself being an $N_f\times N_f$ hermitian matrix, and the potential
$V(Y)$ as yet unspecified, this is a generalized ``external field
problem'' of Random Matrix Theory. The universality of its correlation
functions have been proved in the microscopic large-$N_f$ limit in \cite{PJZ}
for polynomial potentials $V(Y)$.
Surprisingly, a closed solution of the integral eq. (\r{taumatrix})
can be written down for any value of $N_f$, and for any potential 
$V(Y)$ that satisfies suitable convergence criteria \cite{KM3Z,Dijk}. 
After diagonalizing the $Y$-matrix,
$Y= U{\mbox{\rm diag}}(y_1,\ldots, y_{N_f})U^{\dagger}$, one obtains the
standard Jacobian of $\Delta(y)^2$, and one can then make use of the
Harish-Chandra integral to obtain
\beq
\tau(X) ~=~ \int\!\prod dy_i \frac{\Delta(y)}{\Delta(x)} \exp\left[\sum_j
(x_jy_j + V(y_j))\right] ~,
\eeq
where the $x_i$ are the $N_f$ eigenvalues of $X$. If one now introduces the
function
\beq
\phi(x) ~\equiv~ \int\! dy e^{xy + V(y)} ~,
\eeq
as well as the derivatives
\beq
\phi_k(x) ~\equiv~ \frac{\partial^k}{\partial x^k}\phi(x) ~=~ 
\int\! dy y^k e^{xy + V(y)} ~,
\label{phik}
\eeq
one sees that the integral is simply
\beq
\tau(X) ~=~ \frac{\det[\phi_{j-1}(x_i)]}{\Delta(x)} ~.
\label{taudet}
\eeq
This shows that $\tau(X)$ is a 
$\tau$-function of the integrable KP hierarchy.

\noi
The expression (\ref{taudet}) has an uncanny resemblance to the
finite-volume partition function if one identifies $x_i=\mu_i^2$. This
is particularly clear if one considers the form (\ref{Adef}), and
starts with the case $\nu=0$. Using the Bessel function identity
\beq
\frac{d^k}{d(x^2)^k}\left(x^{N_f}I_{N_f}(x)\right) ~=~ \frac{1}{2^k}
x^{N_{f}-k}I_{N_{f}-k}(x) ~,
\eeq
we see that the partition function (\ref{ZUE}) can be written in the form
(\ref{taudet}) if we identify 
\beq
\phi(\mu^2) ~=~ \left(\sqrt{\mu^2}\right)^{N_f}I_{N_{f}}(\sqrt{\mu^2}) ~,
\label{map}
\eeq
and ignore irrelevant overall factors.
The case of non-zero $\nu$ can then be treated by using again the 
flavor-topology duality \cite{Jac}, thus obtaining the $\nu \neq 0$ case
by simply setting $\nu$ of the $N_f$ masses to zero.
In fact, eq. (\r{map}) alone suffices to prove that the effective partition 
function is a $\tau$-function of the integrable KP hierarchy \cite{KM3Z}. 
It is nevertheless interesting to note that in addition an integral
representation actually exists such that the partition function explicitly
can be written in the form (\ref{taumatrix}) \cite{MMS}. This turns out
to correspond to a potential $V(Y) = 1/Y - N_f\ln(Y)$, and an integration
contour for the eigenvalues encircling the origin. (Strictly speaking
this is outside the scope of the hermitian matrix formulation 
(\ref{taumatrix}), so the notion of ``hermiticity'' is here simply taken
to mean ``of flat measure'' -- see ref. \cite{MMS} for a discussion of this
point).

\noi
Knowing that the partition function is a $\tau$-function immediately
implies a number of identities. 
The most general of these is the following set of Hirota equations, which read
\cite{KM3Z}
\beqn
0 &=&(x_a-x_b)\tau(X;p_a,p_b,p_c+1)\tau(X;p_a+1,p_b+1,p_c) \nn\\
 &+& (x_b-x_c)\tau(X;p_a+1,p_b,p_c)\tau(X;p_a,p_b+1,p_c+1) \nn\\
 &+& (x_c-x_a)\tau(X;p_a,p_b+1,p_c)\tau(X;p_a+1,p_b,p_c+1) \ .
\eeqn
Here the $p_i$ denote the multiplicities of the parameters $x_i$ for $i=a,b,c$,
where in our notation $x_i = \mu_i^2$.
Due to the Jacobi identity for determinants yet another identity holds
for $\tau$-functions as given in $e.g.$ eq. (2.43) of ref.\cite{KM3Z}:
\beqn
\tau^{(N_f+2)} (\{x\},x_{N_f+1},x_{N_f+2}) &=&
\frac{1}{(x_{N_f+1}-x_{N_f+2})\tau^{(N_f)} (\{x\})} \ \times \nn\\
&\times& \left[ \tau^{(N_f+1)} (\{x\},x_{N_f+1})
\hat{\tau}^{(N_f+1)} (\{x\},x_{N_f+2}) \ -\ (x_{N_f+1} \leftrightarrow
x_{N_f+2})\right] \ . \nn\\
\eeqn
The upper index indicates the number of parameters of the corresponding 
$\tau$-function and $\hat{\tau}$ means that in the last row the index
of the functions $\phi_k(x)$ in eq. (\r{phik}) has been shifted by $+1$.
This relation looks remarkably similar to our Consistency Condition II
eq. (\r{cond2}). However, in eq. (\r{cond2}) the derivatives, which shift 
the indices of the $\phi_k$ are taken with respect to all variables.
In the derivation of the Consistency Condition II we have used 
properties of the Bessel-functions in only one step, namely in eq. (\r{dI}).
{}The fact that all $\tau$-functions eq. (\r{taudet}) obey the property
\beq
2\mu\partial_\mu \phi_k(x) \ =\ \partial_x \phi_k(x)|_{x=\mu^2}
\ =\ \phi_{k+1}(x)
\eeq
can probably be used to show 
that all $\tau$-functions obey our Consistency Condition II.

\noi
There exists another set of relations for $\tau$-functions of the KP
hierarchy which apparently can be related to our finite-volume partition 
functions. According to ref. \cite{Zabrodin} these relations read
\beq
\det_{1\leq a,b \leq k}\left[ \frac{\tau(\tb +[\xi_a^{-1}]-
[\eta_b^{-1}])}{(\xi_a-\eta_b)\tau(\tb)}   \right] \ =\ 
\prod_{a<b}^k(\xi_a-\xi_b)(\eta_b-\eta_a)
\frac{\tau\left(\tb +\sum_{a=1}^k([\xi_a^{-1}]-
[\eta_a^{-1}])\right)}{\prod_{a,b}^k(\xi_a-\eta_b)\tau(\tb)} \ .
\l{Za}
\eeq
At first sight they look remarkably similar to our Consistency 
Condition I in the 
form of eq. (\r{cond1}). In order to to explain the differences let us 
give the notation of eq. (\r{Za}) from ref. \cite{Zabrodin}. 
The argument of the $\tau$-function $\tb$
stands for all the coupling constants or times in a matrix 
potential $V(\lambda)=\sum_{k=1}^\infty t_k \lambda^k$. 
The bracket [ ] then is a shorthand notation for 
\beq
\tau(\tb \pm[\xi])\ =\ \tau(t_1\pm\xi, t_2\pm\frac{1}{2}\xi^2, 
t_3\pm\frac{1}{3}\xi^3, \ldots) \ .
\eeq
Performing the sum over the additional parameter $[\xi^{-1}]$
in the potential leads to an additinal logarithmic term 
$V(\la)\to V(\la)-\ln(1-\la/\xi)$, 
which resembles an extra ``mass term'' if we could consider this
as an ordinary hermitian Random Matrix Theory in which one has taken 
the microscopic limit. Such an identification is, however, far from
obvious. Moreover, in eq. (\r{Za}) these extra terms occur in pairs
$\xi$ and $\eta$ with {\it opposite} signs and thus
one of them appears as a bosonic ``mass term''.
Due to this difference, apart from other additional factors,
we have not been able to explicitly match our Consistency Condition I 
eq. (\r{cond1})
with eq. (\r{Za}). Moreover, no direct proof has been given in 
ref. \cite{Zabrodin} for the relation eq. (\r{Za}), and we have not been
able to find it elsewhere.

\setcounter{equation}{0}
\section{Summing over topological charges}

\noi
So far our discussion has been restricted to finite-volume partition
functions ${\cal Z}_{\nu}^{(N_{f})}(\{\mu\})$ in sectors of fixed topological
index $\nu$. These partition functions can be thought of as Fourier 
coefficients of the full partition function, which for given vacuum angle 
$\theta$, is given by
\beq
{\cal Z}^{(N_f)}(\theta,\{\mu\}) ~=~ 
\sum_{\nu=-\infty}^{\infty}e^{i\nu\theta}{\cal Z}_{\nu}^{(N_f)}(\{\mu\}) ~,
\eeq
or, in terms of the effective partition function on the coset of
chiral symmetry breaking in this case, 
\beq
{\cal Z}^{(N_f)}(\theta,\{\mu\}) ~=~ \int_{SU(N_{f})}\! dU
\exp\left[V\Sigma {\mbox{\rm Re}}\,[e^{i\theta/N_{f}}{\mbox{\rm Tr}}\,
{\cal M}U^{\dagger}]\right] ~. \label{Zeff}
\eeq

\noi
Contrary to the effective partition functions 
${\cal Z}_{\nu}^{(N_{f})}(\{\mu\})$
in sectors of fixed gauge field topology, the group integral of eq.
(\ref{Zeff}) is not known in closed form for $N_f \geq 3$. We are
not aware of any analogue of the theorems discussed above for the full
partition functions, and in view of the non-linearity of the relations
it seems unlikely that they could be established. Nevertheless, as an 
interesting by-product of the
above analysis we are now able to provide a simple compact formula for
any $k$-point spectral correlation function after having summed
over all topological charges.

\noi
The first observation is that for any observable 
$\langle{\cal O}\rangle_{\nu}$ in the fixed-$\nu$ theory one finds the
same observable in the full theory by summing over $\nu$ with weight
factor $e^{i\nu\theta}{\cal Z}_{\nu}^{(N_f)}(\{\mu\})$:
\beq
\langle\langle{\cal O}\rangle\rangle ~=~ \sum_{\nu=-\infty}^{\infty}
e^{i\nu\theta}{\cal Z}_{\nu}^{(N_f)}(\{\mu\})\ \langle{\cal O}\rangle_{\nu} ~.
\eeq
Next, noting that the $k$-point spectral correlation function by itself
is just an expectation value, also this function can be summed over
topological charges:
\beq
\bar{\rho}_S(\xi_1,\ldots,\xi_k;\theta,\{\mu\}) = 
{\cal Z}^{(N_f)}(\theta;\{\mu\})^{-1}
\sum_{\nu=-\infty}^{\infty}e^{i\nu\theta}{\cal Z}_{\nu}^{(N_f)}(\{\mu\})\ 
\rho_S^{(\nu)}(\xi_1,\ldots,\xi_k;\{\mu\}) ~,\label{spcorr}
\eeq
where we have {\em not} included the zero-mode contributions in the spectral
sums, and where we have already taken the microscopic limit.

\noi
We now make us of the fact that the $k$-point function in a sector of
fixed topological charge $\nu$ can be expressed in terms of a partition
functions with $2k$ additional species as in eq. (\ref{corrft}). 
As a side result of proving the Consistency Condition I in section 2
we have already fixed 
the constant $C_2^{(k)}=(-1)^{k(\nu+[N_f/2])}$. Only the $\nu$-dependence
is important for the summation over topological charges, where it leads
to a shift in the $\theta$-angle:  
\beqn
\bar{\rho}_S(\xi_1,\ldots,\xi_k;\theta,\{\mu\}) &=&
{\cal Z}^{(N_f)}(\theta;\{\mu\})^{-1}
\sum_{\nu=-\infty}^{\infty}e^{i\nu\theta}(-1)^{k(\nu+[N_f/2])}
\prod_i^k\left( |\xi_i|\
\prod_{f=1}^{N_f}(\xi_i^2+\mu_f^2)\right) \ \times\nn\\
&&\ \times\prod_{j<l}^k(\xi_j^2-\xi_l^2)^2 
{\cal Z}_{\nu}^{(N_{f}+2k)}(\{\mu\},\{i\xi_1\},\ldots, \{i\xi_k\})
\nn\\
&=& (-1)^{k[N_f/2]}\prod_{i=1}^k\left(|\xi_i|\prod_{f=1}^{N_f}
(\xi_i^2+\mu_f^2) \right)\prod_{j<l}^k (\xi_j^2-\xi_l^2)^2 \ \times\nn\\
&&\ \times
\frac{{\cal Z}^{(N_f+2k)}(\theta+k\pi;\{\mu\},\{i\xi_1\},\ldots,\{i\xi_k\})}
{{\cal Z}^{(N_f)}(\theta;\{\mu\})} \ \ .
\eeqn 
Due to the periodicity of the angle $\theta$ we need to know the partition
function at either a shifted or unshifted vacuum angle $\theta$ for the 
$k$-point correlation function with $k$ either even or odd
\footnote{In \cite{D99} a factor of $(-1)^{[N_f/2]}$ is missing in the formula
for the density (corresponding here to $k=1$).}.

\setcounter{equation}{0}
\section{Fermions in the adjoint representation}

\noi
With $N_f$ fermions taken in the adjoint representation of the gauge group 
SU($N_c\geq 2)$ the pattern of spontaneous chiral symmetry breaking, 
if it occurs at all, is believed to proceed according to SU$(N_f) \to$
SO$(N_f)$. In the Random Matrix Theory classification this corresponds
to the chiral Symplectic Ensemble chSE. Fermions in the adjoint representation
of the gauge group occur in $e.g.$ supersymmetric gauge theories even without
matter fields, but its interest in the present context stems more from
the fact that staggered fermions in the {\em fundamental} representation and
gauge group SU(2) actually also fall into this universality class away from
the continuum limit. In ref. \cite{SmV} the effective partition function
in the same finite-volume limit as above was written, in a sector of fixed
topological charge $\bar{\nu}= N_c\nu$, in terms of a relatively simple-looking
group integral over the unitary group U($N_f)$:
\beq
{\cal Z}_{\bar{\nu}}^{(N_f)}({\cal M}) ~=~ \int \!dU (\det U)^{-2\bar{\nu}}
\exp\left[\Sigma V{\mbox{\rm Re Tr}}{\cal M}UU^T\right] ~,
\label{Zbeta4}
\eeq
where as before ${\cal M}$ denotes the mass matrix.
This group integral is surprisingly difficult to perform explicitly, and
it has in fact until now only been evaluated for $N_f=2$ fermions of equal 
mass, and any topological charge $\bar{\nu}$ \cite{D0}. 
In general, the equal-mass
partition functions  are much easier to evaluate due to an interesting 
rewriting of the group integral (\ref{Zbeta4}), which is valid for those 
cases \cite{SmV}. We shall here use that form to explicitly evaluate the 
group integral for $N_f=4$ equal-mass fermions in a sector of zero 
topological charge (the derivation extends straightforwardly to any 
topological charge, but we have not considered that extension in detail).

\noi
The partition function (\ref{Zbeta4}) of $N_f$ equal masses has been be 
rewritten by Smilga and Verbaarschot as (for $N_f$ even) \cite{SmV}:
\beq
{\cal Z}_{\bar{\nu}}^{(N_f)}({\cal M}) ~=~ {\mbox{\rm Pf}}(A) ~, \label{Pfaff}
\eeq
where, in our normalization, the $N_f\times N_f$ matrix $A$ has elements
\beq
A_{pq} ~=~ -\frac{i\pi}{2}\int_{-\pi}^{\pi}\frac{d\theta}{2\pi}
\int_{-\pi}^{\pi}\frac{d\phi}{2\pi}\epsilon(\theta-\phi)
e^{i(p\phi+q\theta)}e^{\mu\cos\phi + \mu \cos\theta + i\bar{\nu}(\phi
+\theta)} ~, \label{Aint}
\eeq
and the indices $p$ and $q$ run from $-\frac{N_f}{2}+\frac{1}{2}$ to
$\frac{N_f}{2}-\frac{1}{2}$. Again $\mu \equiv m\Sigma V$ is the (common)
rescaled mass. A for our purposes convenient infinite-sum
representation was also given in ref. \cite{SmV}, based on a Fourier
series expansion of the sign function in eq. (\ref{Aint}):
\beq
A_{pq} ~=~ \sum_{k=-\infty}^{\infty} \frac{1}{2k+1}
I_{\bar{\nu}+p+k+\frac{1}{2}}(\mu) I_{\bar{\nu}+q-k-\frac{1}{2}}(\mu) ~.
\label{Asum}
\eeq
It is with this form of the matrix $A$ that we have managed to evaluate the
partition function (\ref{Zbeta4}) for $N_f=4$ equal masses and
$\bar{\nu}=0$ (this last restriction can readily be lifted). After
some tedious algebra we find (technical details can be found in
Appendix C):
\beq
{\cal Z}_0^{(4)}(\mu) ~=~ \frac{1}{\mu^2}\left[I_1(2\mu)^2 - I_0(2\mu)^2\right]
+ \frac{1}{2\mu^3}I_0(2\mu)\int_0^{2\mu} dt I_0(2\mu) ~. \label{Z4eff}
\eeq
The last integral is explicitly known in terms of a combination of
Struve and Bessel functions, but we leave the result like this in order to
facilitate a comparison to be discussed below.

\noi
The reason for our interest in the partition function (\ref{Zbeta4}) is
a general relation derived in the third of ref. \cite{AD}, which expresses
the microscopic spectral density of the Dirac operator for this case in 
terms of the partition function itself and the partition 
function with 4 additional fermion species of imaginary (degenerate)
masses:  
\beq
\rho_S^{(\nu)}(\xi;\{\mu\}) ~=~ C_4~ \xi^3~(\xi^2+\mu^2)^4~
\frac{{\cal Z}^{(N_{f}+4)}_\nu(\{\mu\},\{i\xi\})}
{{\cal Z}^{(N_{f})}_\nu(\{\mu\})} ~.
\label{symrho}
\eeq
The normalization coefficient $C_4$ can be fixed as soon as one settles
on the normalization of the partition functions. The general formula
(\ref{symrho}) as derived in ref. \cite{AD} in the Random Matrix Theory
formulation (and the partition functions involved were therefore those
of Random Matrix Theory, too). But in the microscopic limit
these coincide, modulo uninteresting mass-independent normalization factors,
with the field theory partition functions (\ref{Zbeta4}), thus giving
explicitly the microscopic spectral density in terms of the field theory
partition functions, as indicated. Although this general formula may
provide a simple way of deriving the massive double-microscopic spectral
density for this universality class, it has not yet been tested due
to the lack of a simple analytical expression for the partition functions
(\ref{Zbeta4}). Now, with the analytical result (\ref{Z4eff}) we can for
the first time check the formula, since for the {\em quenched} case
(formally defined by taking $N_f$ to zero) the partition function itself
becomes an uninteresting constant (which we take to be unity), while
the partition function in the numerator of eq. (\ref{symrho}) is that
of just four fermions (of imaginary and degenerate masses). 

\noi
Using well-known relations between Bessel functions and modified Bessel
functions, we thus derive the quenched microscopic spectral density
of the Dirac operator for this case:
\beqn
\rho_S^{(0)}(\xi) &=& C_4 \xi^3 {\cal Z}_0^{(4)}(\{i\xi\})\cr
&=& C_4\left\{\xi[J_0(2\xi)^2 + J_1(2\xi)^2]
- \frac{1}{2}J_0(2\xi)\int_0^{2\xi}\! dt J_0(t)\right\} ~.
\eeqn
If we next impose the matching condition $\rho_s(\xi\to\infty) = 1/\pi$, 
then the overall constant in front is fixed to $C_4=1$. This finally gives
\beq
\rho_S^{(0)}(\xi) ~=~ \xi[J_0(2\xi)^2 + J_1(2\xi)^2]
- \frac{1}{2}J_0(2\xi)\int_0^{2\xi}\! dt J_0(t) ~, 
\eeq
which agrees exactly with the result obtained directly from Random Matrix
Theory \cite{F}. 

\noi
In a derivation of the formula (\ref{symrho}) directly from a partially
quenched chiral Lagrangian, the four-fold mass-degenerate additional 
fermion species should come out from coset of the supergroup chiral
symmetry breaking of that case. The fact that one is led to {\em four}
additional quarks is a somewhat surprising feature, as in a very naive
counting one could have expected {\em two}: one from the additional
quenched quark, and one from its supersymmetric partner. It is therefore
very comforting to see that the recent explicit computation of the
partially quenched effective lagrangian by Toublan and Verbaarschot
\cite{TV} in this case precisely leads to four additional species in total. 
By compactifying variables after taking the discontinuity that gives the
spectral density it should therefore now be possible to derive the
formula (\ref{symrho}) directly from the effective Lagrangian, following
the steps of the first of ref. \cite{OTV}.

\setcounter{equation}{0}
\section{Conclusions}

\noi
Our main purpose here has been to show that the surprising relations
among the effective partition functions relevant for describing the
microscopic Dirac operator spectrum can be derived directly, without
recourse to the Random Matrix Theory formulation. We have noted that
these identities most likely have as their origin the fact that the
effective partition functions of the chUE universality class are 
$\tau$-functions of an integrable KP hierarchy. We believe
that one of these identities, here called Consistency Condition II,
holds in general for all these $\tau$-functions. 

\noi
As a by-product of our analysis, we have computed the $\nu$-dependent
normalization factor of the $k$-point spectral correlation function
of the same universality class. This has allowed us to perform the
sum over topological charges $\nu$ explicitly, and express the $k$-point
function of the full theory entirely in terms of the full effective
partition functions, without the restriction to fixed topological
charge.

\noi
We have noted a series of relations between the effective
finite-volume partition functions for QCD$_3$-like theories and
QCD$_4$-like theories. These relations translate into surprising
relations between the external $U(N_f)$ field problem and the Harish-Chandra
integral for unitary groups.

\noi
Finally, we have considered an analogous formula for the
microscopic spectral density of the chSE universality class, which
expresses this spectral density in terms of the effective field theory
partition function with four additional (imaginary-mass) fermion
species. We have explicitly shown that this formula yields the same 
analytical result as the Random Matrix Theory approach in the quenched
case of $N_f=0$. Considering the analytical difficulties in extending
the corresponding Random Matrix Theory calculation to the case of
massive fermions, this may provide the most economical way of deriving
all the microscopic spectral correlators of that universality class.

\vspace{1cm}
\noi
{\sc Acknowledgment:}\\
We thank V. Kazakov and I. Kostov for pointing out the possible relation 
between our consistency conditions and Hirota equations. 
Furthermore G.A. would like to thank T. Guhr and H. Kohler for helpful 
discussions. This work was supported
in part by EU TMR grant no. ERBFMRXCT97-0122.

\appendix

\setcounter{equation}{0}
\section{Some lemmas for Consistency Condition I} \label{A}

\underline{{\sc Lemma 1}:}~
Let $A\left(\{\xi\},\{\eta\}\right)$ be the following $2k\times 2k$ matrix 
\beq
\left(A(\{\xi\},\{\eta\})\right)_{ij} \ =\ 
\left\{ \ba \xi_i^{j-1}I_{\nu+j-1}(\xi_i)& & i=1,\ldots,k \\
\eta_i^{j-1}I_{\nu+j-1}(\eta_i)& & i=k+1,\ldots,2k \ea \right. \ \forall \ j\ .
\eeq
Then the following statement holds:
\beq
\det A(\{\xi\},\{\eta\}) \ =\ (-1)^{k(k-1)/2}\left| \ba B(\xi)& & C(\xi)\\
                                             B(\eta)& & C(\eta) \ea \right|
\ \prod_{i=1}^k I_{\nu}(\xi_i)I_{\nu}(\eta_i)
\l{la1}
\eeq
where the two $k\times k$ matrices $B$ and $C$ are given by
\beq
B(\xi)_{ij} \ =\ (\xi_i^2)^{j-1} \ ,\ \ \ \ C(\xi)_{ij} \ =\ 
(\xi_i^2)^{j-1} \xi_i\vec{\partial}_{\xi_i}\ ,\ \ i,j=1,\ldots,k \ \ .
\l{BC}
\eeq
In other words $B$ is the matrix inside the Vandermonde with squared
arguments whereas $C$ contains in addition 
the power counting operator in each row, which acts on Bessel functions
outside the determinant only.

\noi
Before proving the lemma let us add a remark. The above statement is not
specific for a matrix with variables split into 2 equal groups, as we will
see in the proof below. Furthermore eq. (\r{la1}) trivially extends to matrices
$A$ of odd size, where one additional column has to be added in the matrix $B$.
We have just presented the statement in the precise form as we need it in
the proof of theorem \r{cond1m}.

\noi
\underline{{\sc Proof}:}~Let us first for simplicity set
\beq
x_i \ \equiv \ \xi_i\ ,\ \ x_{i+k}\ \equiv \ \eta_i \ \ 
\mbox{for}\ \ i=1,\ldots,k \ .
\l{xxieta}
\eeq
We will in a first step use the property of the Bessel functions
\beq
xI_{n+1}(x) \ + \ 2n I_n(x) \ =\ xI_{n-1} 
\eeq
to reduce all indices of the Bessel functions in det$A(\{x\})$ down to $\nu$ or
$\nu+1$. Starting with the last column we can add $2(\nu+2k-2)$ times the last
but one column to it to reduce the index of the Bessel function by 2:
\beq
x_i^{2k-1}I_{\nu+2k-1}(x_i) \ + \ 2(\nu+2k-2)x_i^{2k-2} I_{\nu+2k-2}(x_i) \ =\ 
x_i^{2k-1}I_{\nu+2k-3}(x_i)\ . 
\l{reduce}
\eeq
We proceed similarly with the last but one column and go down till the third
column. All indices of Bessel functions have been reduced by 2 except in the
first 2 columns. We then start again with the last column reducing the index
by 2 (eq. (\r{reduce}) for $\nu\to\nu-2$) and continue down to the fifth
column. So after going down through all columns $k-1$ times always starting
from the right we have achieved 
\beq
\det A(\{x\}) \ =\ \det_{1\leq i,j \leq 2k}
\left( x_i^{j-1} I_{\nu+\frac{1+(-1)^j}{2}}(x_i)\right) \ .
\eeq

\noi
In a second step we write all functions $I_{\nu+1}(x)$ which appear in every 
second column as a derivative $\partial_x I_\nu(x)$ so that we can factor out
all Bessel functions to the right of the determinant. Using eq. (\r{dI})
for $n=0$ we have 
\beq
x^{2l+1}I_{\nu+1}(x) \ =\ x^{2l+1}\partial_x I_{\nu}(x) \ -\ \nu x^{2l}I_\nu(x)
\ .
\l{dInu}
\eeq
Inserting this expression into every second column we can again compensate the 
last term in eq. (\r{dInu}) by adding $\nu$ times the column to the left.
We thus have
\beq
\det A(\{x\})\ =\ \det \tilde{A}(\{x\}) \prod_{i=1}^k I_\nu(x_i) \ ,
\eeq
where
\beq
(\tilde{A}(\{x\}))_{ij} \ =\ \left\{ \ba x_i^{j-1}& & j=1,3,\ldots,2k-1 \\
x_i^{j-1}\vec{\partial}_{x_i}& & j=2,4,\ldots,2k \ea \right. \ \forall 
\ i=1,\ldots,2k\ .
\eeq
Here we have pulled out all common factors $I_\nu(x_i)$ out of each row $i$
and again the derivatives only act on these.
After reordering those columns with even powers to the left and those with
odd powers times a derivative to the right, which results into a factor
$(-1)^{k(k-1)/2}$, we obtain eq. (\r{la1}) when renaming back the
variables from eq. (\r{xxieta}).

\noi
\underline{{\sc Lemma 2}:}~Let
\beq
(G(i:\ga))_{jl} \ \equiv\ \left\{
\ba \frac{1}{\eta_{\ga(j)}^2-\xi_l^2}\xi_l\vec{\partial}_{\xi_l}& 
& l=1,\ldots,i\\
\frac{1}{\eta_{\ga(j)}^2-\xi_l^2} & & l=i+1,\ldots,k \ea \right. 
\forall\ j=1,\ldots,k \ \ ,
\l{Gidef}
\eeq
where $\ga$ is a permutation of the indices $1,\ldots,k$ with the ordering
$\ga(1)<\ldots<\ga(i)$ and $\ga(i+1)<\ldots<\ga(k)$ for $1\leq i\leq k$ fixed.
It then holds
\beqn
\det G(i;\ga) &=&
(-1)^{\frac{k(k-1)}{2}}\prod_{a,b=1}^k\frac{1}{(\eta_a^2-\xi_b^2)} 
\sum_{\bar{\ga}}(-1)^{\bar{\ga}}
\Delta(\xi_{\bar{\ga}(1)}^2,\dots,\xi_{\bar{\ga}(k-i)}^2,
\eta_{\ga(1)}^2,\dots,\eta_{\ga(i)}^2) \times \nn\\
&&\times \ 
\Delta(\xi_{\bar{\ga}(k-i+1)}^2,\dots,\xi_{\bar{\ga}(k)}^2,
\eta_{\ga(i+1)}^2,\dots,\eta_{\ga(k)}^2) 
\prod_{j=k-i+1}^k \xi_{\bar{\ga}}(j)\partial_{\xi_{\bar{\ga}(j)}} \ ,
\l{la2}
\eeqn
where $\bar{\ga}$ is a permutation as $\ga$ of the indices of the $\xi$'s.

\noi
\underline{{\sc Proof}:}~We will prove Lemma 2 by doing a 
Laplace expansion with respect to the first 
$i$ columns into $i$ times $(k-i)$ dimensional determinants. Each of the 
subdeterminants can then be evaluated by using Cauchy's Lemma eq. (\r{Cauchy})
for squared variables, which reads
\beq
\det_{1\leq a,b\leq k}\left[\frac{1}{\eta_b^2-\xi_a^2}\right] \ =\ 
(-1)^{\frac{k(k-1)}{2}}\frac{\prod_{a<b}^{k}(\xi_a^2-\xi_b^2)
(\eta_a^2-\eta_b^2)}{\prod_{a,b=1}^k(\eta_a^2-\xi_b^2)} \ .
\l{Cauchy2}
\eeq
We thus obtain for the first upper block
\beq
\det_{1\leq j,l\leq i}\left[ 
\frac{1}{\eta_{\ga(j)}^2-\xi_l^2}\xi_l\vec{\partial}_{\xi_l}\right] \ =\ 
(-1)^{\frac{i(i-1)}{2}}\frac{\prod_{a<b}^{i}(\xi_a^2-\xi_b^2)
(\eta_{\ga(a)}^2-\eta_{\ga(b)}^2)}{\prod_{a,b=1}^i(\eta_{\ga(a)}^2-\xi_b^2)}
\prod_{j=1}^i \xi_j\partial_{\xi_j} \ ,
\eeq
and similarly for the lower blocks, where we can directly use 
eq. (\r{Cauchy2}). Taking all the permutations $\bar{\ga}$ according to the 
Laplace expansion we arrive after a few steps at
\beqn
\det G(i;\ga) &=&
(-1)^{\frac{k(k-1)}{2}}(-1)^{i(i-k)}
\prod_{a,b=1}^k\frac{1}{(\eta_a^2-\xi_b^2)} 
\sum_{\bar{\ga}}(-1)^{\bar{\ga}}
\Delta(\xi_{\bar{\ga}(1)}^2,\dots,\xi_{\bar{\ga}(i)}^2,
\eta_{\ga(i+1)}^2,\dots,\eta_{\ga(k)}^2) \times \nn\\
&&\times \ 
\Delta(\xi_{\bar{\ga}(i+1)}^2,\dots,\xi_{\bar{\ga}(k)}^2,
\eta_{\ga(1)}^2,\dots,\eta_{\ga(i)}^2) 
\prod_{j=1}^i \xi_{\bar{\ga}}(j)\partial_{\xi_{\bar{\ga}(j)}} \ ,
\l{la2shift}
\eeqn
where the sign of the permutation is defined by
\beq
(-1)^{\bar{\ga}}\ =\ (-1)^{\sum_{j=1}^i(j+\bar{\ga}(j))} 
\ =\ (-1)^{\sum_{j=i+1}^k(j+\bar{\ga}(j))} \ . 
\l{signperm}
\eeq
To obtain the final form of eq. (\r{la2}) we still have to perform a cyclic 
shift by $-i$ places in the indices permuted by $\bar{\ga}$. Using 
the second form of eq. (\r{signperm}) we obtain
\beqn
(-1)^{\bar{\ga}\ +\ i(i-k)} 
&\rightarrow& (-1)^{ \frac{k(k+1)}{2}\ -\ \frac{i(i+1)}{2}
\ +\ \sum_{j=1}^{k-i}\bar{\ga}(j) \ +\ i(i-k)}
\nn\\
&=&(-1)^{\sum_{j=1}^{k-i}(j+\bar{\ga}(j))} \ ,
\eeqn
which leads us from eq. (\r{la2shift}) to eq. (\r{la2}).

\setcounter{equation}{0}
\section{Some lemmas for Consistency Condition II} \label{B}

\underline{{\sc Lemma 3}:}~Let $A$ be 
an $n\times n$ matrix with elements $a_{kl}$ and $A_j$
be the same matrix with the $j$-th row shifted by one unit to the left:
$(A_j)_{kl}=a_{kl}$ for $k\neq j$ and $(A_j)_{jl}=a_{jl+1}$.
Expanding the determinant of $A$ with respect to the last column we have
\beq
\det A\ =\ \sum_{l=1}^n (-1)^{n+l} a_{ln} \det A^*_{ln}
\eeq
where $A^*_{ln}$ is the algebraic complement of $a_{ln}$. It then holds
\beq
\sum_{j=1}^n \det A_j \ =\ \sum_{j=1}^n (-1)^{n+j} a_{jn+1} \det A^*_{jn}
\l{Lemma3}
\eeq

\noi
\underline{{\sc Proof}:}~
We have to show that when summing over the determinants of the shifted matrices
$A_j$ and expanding each of them with respect to the last column only 
determinants of the algebraic complement of $A$ remain, all the other terms 
containing determinants of shifted matrices cancel.

\noi
Starting from the definition we obtain after rearranging terms
\beq
\sum_{j=1}^n \det A_j \ = \sum_{j=1}^n (-1)^{n+j} a_{jn+1} \det A^*_{jn}
  \ +\ \sum_{i=1}^n (-1)^{n+i} a_{in}\sum_{j=1\ ;\ j\neq i}^n  
\det (A_j)^*_{in}\ .
\l{eqla3}
\eeq
Here we have used the same notation for the shifted matrices and applied
the fact that $(A_j)^*_{jn}=A^*_{jn}$. We will now show by induction that
the double sum in eq. (\r{eqla3}) vanishes. For $n=2$ this holds trivially.
Inside the double sum we sum over shifted matrices of size $n-1$ so we can 
apply the statement eq. (\r{Lemma3}) for $n-1$. Introducing the 
$(n-2)\times(n-2)$ submatrix $A^*_{jn-1;in}$ of $A$, where now the last
2 columns and rows $i$ and $j$ are missing we are left with
\beq
  0\ =\ \sum_{i=1}^n(-1)^{n+i}a_{in}\left[
     \sum_{j=1}^{i-1} (-1)^{n-1+j} a_{jn} \det A^*_{jn-1;in} 
     \ +\ \sum_{j=i+1}^{n} (-1)^{n+j} a_{jn} \det A^*_{jn-1;in} \right]
\eeq
where the split is due to the missing row when expanding twice. It then follows
from the obvious symmetry $ A^*_{in-1;jn}= A^*_{jn-1;in}$ that the 
two sums inside the bracket cancel.

\noi
\underline{{\sc Lemma 4}:}~
Let $A$ and $A^*_{ln}$ be the same as in Lemma 3 and let $A^L_{k}$ be 
the matrix $A$ with the first $L$ entries vanishing in the $k$-th
row: $(A^L_{k})_{ij}=0$ for $i=1,\ldots,L,\ j=k$ and $a_{ij}$ else-wise.
Then it holds for any fixed $L$
\beq
\sum_{k=1}^n \det A_k^L \ =\ (n-L) \det A \ .
\l{Lemma4}
\eeq

\noi
\underline{{\sc Proof}:}~
Let us first state the trivial cases apart from
$L=0,n$. For $L=1$ the determinants just
differ by the first column and can thus be added up, giving $(n-1)\det A$
after pulling out a common factor. For $L=n-1$ each $A_k^L$ can be
expanded with respect to the row $k$ which then just gives $\det A$.

\noi
We will now proceed by induction. Due to the above remarks the cases $n=2,3$
are trivial. If we expand the determinants on the left hand side of 
eq. (\r{Lemma4}) with respect to the last column we obtain
\beqn
\sum_{k=1}^n \det A_k^L &=& 
        \sum_{k=1}^n \sum_{i=1;i\neq k}^n(-1)^{n+i} a_{in}\det (A_k^L)^*_{in}
 \ + \ \sum_{i=1}^n(-1)^{n+i}a_{in}\det A^*_{in}  \nn\\
&=&  \sum_{i=1}^n(-1)^{n+i}a_{in}(n-1-L)\det A^*_{in}
 \ + \ \det A \ \ .
\l{eqLemma4}
\eeqn
In the first step we have used that $(A_i^L)^*_{in}=A^*_{in}$
and in the second step we have employed 
induction for $n-1$. The sum in the last
line then gives $(n-1-L)\det A$ and adds up to the statement eq. (\r{Lemma4}).

\noi
\underline{{\sc Lemma 5}:}~
Let $A$ be a nonsingular $(n-2)\times n$ matrix and $b,c,\xi$ and $\eta$
be $n$-vectors. Then the following property holds for determinants of
$n\times n$ matrices which differ by the last two rows or columns:
\beq
\left|\ba &A&\\&b&\\&\xi& \ea \right| \left|\ba &A&\\&c&\\&\eta& \ea \right|
\ -\ \left|\ba &A&\\&b&\\&\eta& \ea \right|\left|\ba &A&\\&c&\\&\xi&\ea\right|
\ =\ \left|\ba &A&\\&b&\\&c&\ea \right|\left|\ba &A&\\&\xi&\\&\eta& \ea \right|
\ . \l{Lemma5}
\eeq

\noi
\underline{{\sc Proof}:}~
Using properties of determinants $\det (BC)=\det B\det C$ and 
$\det A=\det A^T$ the statement eq. (\r{Lemma5}) is equivalent to
\beq
\left|\ba AA^T&Ac&A\eta\\ bA^T&bc&b\eta\\ \xi A^T&\xi c&\xi\eta \ea \right|
\ -\ 
\left|\ba AA^T&Ac&A\xi\\ bA^T&bc&b\xi\\ \eta A^T&\eta c&\eta\xi \ea \right|
\ =\ 
\left|\ba AA^T&A\xi&A\eta\\ bA^T&b\xi&b\eta\\ cA^T&c\xi &c\eta \ea \right|
\ .\l{eqLemma5}
\eeq
We will now use extensively that determinants that differ by one row can be
added. We obtain for the left hand side
\beqn
\mbox{l.h.s.} &=&
\left|\ba AA^T&Ac&A\eta\\ 0&bc&0\\ \xi A^T&\xi c&\xi\eta \ea \right|
+ \left|\ba AA^T&Ac&A\eta\\ bA^T&0&b\eta\\ \xi A^T&\xi c&\xi\eta \ea \right| 
-\left|\ba AA^T&Ac&A\xi\\ 0&bc&0\\ \eta A^T&\eta c&\eta\xi \ea \right|
-\left|\ba AA^T&Ac&A\xi\\ bA^T&0&b\xi\\ \eta A^T&\eta c&\eta\xi \ea \right|
\nn\\
 &=&
\left|\ba AA^T&Ac&A\eta\\ bA^T&0&b\eta\\ 0&0&\xi\eta \ea \right| 
+\left|\ba AA^T&Ac&A\eta\\ bA^T&0&b\eta\\ \xi A^T&\xi c&0 \ea \right| 
-\left|\ba AA^T&Ac&A\xi\\ bA^T&0&b\xi\\ 0&0&\eta\xi \ea \right|
-\left|\ba AA^T&Ac&A\xi\\ bA^T&0&b\xi\\ \eta A^T&\eta c&0 \ea \right|
\nn\\
&=&
\left|\ba AA^T&A\xi&Ac\\ bA^T&b\xi&0\\ \eta A^T&0&0 \ea \right|
-\left|\ba AA^T&A\eta&Ac\\ bA^T&b\eta&0\\ \xi A^T&0&0 \ea \right| 
+\left|\ba AA^T&Ac&A\eta\\ bA^T&0&b\eta\\ 0&c\xi&0 \ea \right| 
+\left|\ba AA^T&A\xi&Ac\\ bA^T&b\xi&0\\ 0&0&c\eta  \ea \right|
\eeqn
In the first line the 1. and 3. term cancel after expanding with respect to 
the last but one row, the same happens in the second line 
expanding the last row. The last line gives the right hand side of 
eq. (\r{eqLemma5}) expanded with respect to the last row if we can show that
\beq
\left|\ba AA^T&A\xi&A\eta\\ bA^T&b\xi&b\eta\\ cA^T&0&0 \ea \right| 
\ =\ 
\left|\ba AA^T&A\xi&Ac\\ bA^T&b\xi&0\\ \eta A^T&0&0 \ea \right|
-\left|\ba AA^T&A\eta&Ac\\ bA^T&b\eta&0\\ \xi A^T&0&0 \ea \right| \ . 
\l{eqLemma5.2}
\eeq
Performing similar steps as before this can be shown to be
equivalent to
\beq
\left|\ba AA^T&A\xi&A\eta\\ bA^T&0&0\\ cA^T&0&0 \ea \right| 
\ =\ 
\left|\ba AA^T&A\xi&Ac\\ bA^T&0&0\\ \eta A^T&0&0 \ea \right| 
-\left|\ba AA^T&A\eta&Ac\\ bA^T&0&0\\ \xi A^T&0&0 \ea \right| \ .
\l{eqLemma5.3}
\eeq
We will now apply the property 
$\det \left(\ba B& &C\\ D& &E\\\ea \right)=\det B\det(E-DB^{-1}C)$
for matrices where $B,E$ are quadratic and $B$ is nonsingular.
Having $E=0_{2\times2}$ 
and $B=AA^T$ nonsingular in the upper left corner eq. (\r{eqLemma5.3})
is equivalent to
\beqn
\det\left( \left( \ba &bA^T&\\&cA^T&\ea\right)\left(AA^T\right)^{-1}
(A\xi\ A\eta)\right) 
&=& 
\det\left( \left( \ba &bA^T&\\&\eta A^T& \ea\right)\left(AA^T\right)^{-1}
(A\xi\ Ac)\right)\nn\\
&-&\det\left( \left( \ba &bA^T&\\&\xi A^T&\ea\right)\left(AA^T\right)^{-1}
(A\eta\ Ac)\right)\nn
\eeqn
and thus to
\beq
\det\left( (A\xi\ A\eta)\left( \ba &bA^T&\\&cA^T&\ea\right)\right) \nn\\
\ =\ \det\left( (A\xi\ Ac)\left( \ba &bA^T&\\&\eta A^T&\ea\right)\right)
-\det\left( (A\eta\ Ac)\left( \ba &bA^T&\\&\xi A^T&\ea\right)\right) \ ,
\eeq
which can be seen to hold by writing out the $2\times2$ determinants.

\setcounter{equation}{0}
\section{The partition function for $N_f=4$ adjoint rep. fermions} 

We shall here give some technical details on the derivation of eq. 
(\ref{Z4eff}) in the main text. Our starting point is the general relation
(\ref{Pfaff}) with, for equal masses, the matrix $A$ given by eq. (\ref{Asum}).
The first observation is that the matrix $A$, apart from being antisymmetric,
also has a mirror symmetry along the line perpendicular to (the conventionally
defined) diagonal. In other words:
\beq
A_{-\frac{3}{2},-\frac{1}{2}} ~=~ A_{\frac{1}{2},\frac{3}{2}} ~~~~,~~~~~~~~
A_{-\frac{3}{2},\frac{1}{2}} ~=~ A_{-\frac{1}{2},\frac{3}{2}} ~.
\label{mirror}
\eeq
This means that we only have to evaluate 4 independent matrix elements.
We choose these to be $a_{12} \equiv A_{-\frac{3}{2},-\frac{1}{2}}, 
a_{13} \equiv A_{-\frac{3}{2},\frac{1}{2}}, a_{14} \equiv A_{-\frac{3}{2},
\frac{3}{2}}$ and $a_{23} \equiv A_{-\frac{1}{2},\frac{1}{2}}$.
(To avoid the cumbersome indices we have changed the notation from matrix $A$
to matrix $a$ as indicated).

\noi
Our trick is to first Taylor expand all relevant expressions by means
of the Taylor expansion of Bessel functions,
\beq
I_b(x) ~=~ \left(\frac{1}{2}x\right)^b
\sum_{k=0}^{\infty}\frac{(\frac{1}{2}x)^{2k}}{k!\Gamma(b+k+1)} ~,
\eeq
and then re-express the result for each matrix element as far as possible 
in terms of Bessel functions again, using the same Taylor expansion. The first
steps of this procedure give
\beqn  
a_{12} &=& \sum_{n=0}^{\infty}\frac{\mu^{2(n+1)}}{(n!)^2(2n+3)(n^2+3n+2)}\cr
&=& \frac{1}{\mu}\sum_{n=0}^{\infty}\frac{\mu^{2n+3}}{n!(n+2)!(2n+3)}\cr
&=& \frac{1}{\mu}\sum_{n=0}^{\infty}\int_0^{\mu}\! dt 
\frac{t^{2n+2}}{n!(n+2)!}\cr
&=& \frac{1}{2\mu}\int_0^{2\mu}\! dt I_2(t) \cr
&=& \frac{1}{\mu}I_1(2\mu) - \frac{1}{2\mu}\int_0^{2\mu}\! dt I_0(t) ~,
\eeqn
after using a simple Bessel function identity. We next evaluate $a_{23}$:
\beqn
a_{23} &=& \sum_{n=0}^{\infty}\frac{\mu^{2n}}{(n!)^2(2n+1)}\cr
&=& \frac{1}{\mu}\sum_{n=0}^{\infty}\int_0^{\mu}\! dt \frac{t^{2n}}{(n!)^2}\cr
&=&  \frac{1}{2\mu}\int_0^{2\mu}\! dt I_0(t) ~,
\eeqn
which shows that we have the relation
\beq
a_{12} + a_{23} ~=~ \frac{1}{\mu}I_1(\mu) ~.\label{a12a23}
\eeq
 
\noi
Next, we evaluate $a_{13}$ using the same procedure:
\beqn
a_{13} &=& \sum_{n=0}^{\infty}\frac{(2n+2)\mu^{2n+1}}{((n+1)!)^2(2n+3)}\cr
&=&\frac{1}{\mu}\left[\sum_{n=0}^{\infty}\frac{\mu^{2(n+1)}}{((n+1)!)^2}
-\sum_{n=0}^{\infty}\frac{\mu^{2(n+1)}}{((n+1)!)^2(2n+3)}\right]\cr
&=& \frac{1}{\mu}\left[\sum_{n=0}^{\infty}\frac{\mu^{2n}}{((n)!)^2} - \left(1 
+\sum_{n=0}^{\infty}\frac{\mu^{2(n+1)}}{((n+1)!)^2(2n+3)}\right)\right]\cr
&=& \frac{1}{\mu}\left[I_0(2\mu) -  
\sum_{n=0}^{\infty}\frac{\mu^{2n}}{(n!)^2(2n+1)}\right]\cr 
&=& \frac{1}{\mu}\left[I_0(2\mu) -  a_{23} \right] ~,\label{a13a23}
\eeqn
which shows that also $a_{13}$ and $a_{23}$ are simply related. Using
the Taylor expansion of $a_{14}$,
\beq
a_{14} ~=~ \sum_{n=0}^{\infty} \frac{(6n+1)\mu^{2n}}{(n!)^2(2n+3)(2n+1)} ~,
\eeq
we find that the following identity holds:
\beq
a_{14} - a_{23} + \frac{1}{\mu} a_{13} ~=~ 2a_{12} ~.\label{many}
\eeq

\noi
{\underline{\sc Proof:}}~ We simply note that the Taylor series for the matrix
elements $a_{14}, a_{23}$ and $a_{13}$ combine in a simple way:
\beqn
&&\sum_{n=0}^{\infty}\left[\frac{(6n+1)\mu^{2n}}{(n!)^2(2n+3)(2n+1)} -
\frac{\mu^{2n}}{(n!)^2(2n+1)} + \frac{1}{\mu}\frac{(2n+2)\mu^{2n+1}}
{((n+1)!)^2(2n+3)} \right]\cr
&=&\sum_{n=0}^{\infty}\frac{\mu^{2n}}{(n!)^2}\left[\frac{6n+1}
{(2n+3)(2n+1)} -
\frac{1}{(2n+1)} + \frac{2}
{(n+1)(2n+3)}\right] \cr
&=& \sum_{n=0}^{\infty}\frac{\mu^{2n}}{(n!)^2}\frac{4n^2+6n}{(2n+3)(2n+1)
(n+1)}\cr
&=& \sum_{n=0}^{\infty}\frac{2\mu^{2(n+1)}}{(n!)^2(n+1)(2n+3)(n+2)}\cr
&=& 2a_{12} ~,
\eeqn
which proves the identity eq. (\r{many}).

\noi
We are now ready to evaluate the Pfaffian of eq. (\ref{Pfaff}), using the
fact the a Pfaffian of a $4\times4$ antisymmetric matrix is
\beqn
{\mbox{\rm Pf}}(A) &=& a_{12}a_{34} - a_{13}a_{24} + a_{14}a_{23} \cr
  &=& a_{12}^2 - a_{13}^2 + a_{14}a_{23} ~,
\eeqn
where in the second line we have used our additional symmetry (\ref{mirror}).

\noi
It is handy to evaluate instead $\mu^3{\mbox{\rm Pf}}(A)$. We then get
successively, using relations (\ref{a12a23}), (\ref{a13a23}) and
(\ref{many}):
\beqn
\mu^3{\mbox{\rm Pf}}(A) &=& \mu^3[a_{12}^2 - a_{13}^2 + a_{14}a_{23}]\cr
&=& \mu I_1(2\mu)^2 + \mu^3a_{23}^2 - 2\mu^2 I_1(2\mu)a_{23} - \mu I_0(2\mu)^2
\cr &~& + \mu I_0(2\mu)a_{23} + \mu^2a_{23}a_{13} + a_{14}a_{23}\mu^3 \cr
&=& \mu[I_1(2\mu)^2 - I_0(2\mu)^2] + \mu I_0(2\mu)a_{23} \cr
&~& + a_{23}[\mu^3a_{23} - 2\mu^2 I_1(2\mu) + \mu^2a_{13} + \mu^3a_{14}]\cr
&=& \mu[I_1(2\mu)^2 - I_0(2\mu)^2] + \mu I_0(2\mu)a_{23} \cr
&~& + a_{23}[\mu^3a_{23} - 2\mu^2(\mu[a_{12}+a_{23}]) + \mu^2a_{13} + 
\mu^3a_{14}]\cr
&=& \mu[I_1(2\mu)^2 - I_0(2\mu)^2] + \mu I_0(2\mu)a_{23} \cr
&~& + \mu^3a_{23}[a_{14} - 2a_{12} - a_{23} + \frac{1}{\mu}a_{13}] \cr
&=& \mu[I_1(2\mu)^2 - I_0(2\mu)^2] + \mu I_0(2\mu)a_{23} \cr
&=& \mu[I_1(2\mu)^2 - I_0(2\mu)^2] + \frac{1}{2}I_0(2\mu)\int_0^{2\mu}\!
dt I_0(t) ~.
\eeqn
This is the result quoted in the main text.


\begin{thebibliography}{X}

\bibitem{SV}E.V. Shuryak and J.J.M. Verbaarschot, Nucl. Phys. {\bf A560}
(1993) 306. 

\bibitem{V}J.J.M. Verbaarschot and I. Zahed, Phys. Rev. Lett. {\bf 70} 
(1993) 3852; Phys. Rev. Lett. {\bf 73} (1994) 2288. 
\newline J.J.M. Verbaarschot,
Phys. Lett. {\bf B329} (1994) 351; Nucl. Phys. {\bf B426} (1994) 559.

\bibitem{ADMN}G. Akemann, P.H. Damgaard, U. Magnea and S. Nishigaki,
Nucl. Phys. {\bf B487} (1997) 721.\newline 
P.H. Damgaard and S.M. Nishigaki, Nucl. Phys. {\bf B518} (1998)
495.\newline
M.K. \c{S}ener and J.J.M. Verbaarschot, Phys. Rev. Lett. {\bf 81}
(1998) 248.   


\bibitem{AD}P.H. Damgaard, Phys. Lett. {\bf B424} (1998)322.\newline
G. Akemann and P.H. Damgaard, Nucl. Phys. {\bf B519} (1998) 682;
Phys. Lett. {\bf B432} (1998) 390.

\bibitem{OTV}J.C. Osborn, D. Toublan and J.J.M. Verbaarschot, Nucl. Phys. 
{\bf B540} (1999) 317.
\newline P.H. Damgaard, J.C. Osborn, D. Toublan and J.J.M. Verbaarschot,
Nucl. Phys. {\bf B547} (1999) 305.  

\bibitem{TV}D. Toublan and J.J.M. Verbaarschot, hep-th/9904199.

\bibitem{HV}M.A. Halasz and J.J.M. Verbaarschot, Phys. Rev {\bf D52} (1995)
2563.

\bibitem{Dtalk}P.H. Damgaard, hep-th/9807026. 

\bibitem{MM}G. Mahoux and M.L. Mehta, J. Phys. {\bf I} (France) (1991) 1093.

\bibitem{Witten}E. Witten, Nucl.\ Phys.\ {\bf B340} (1990) 281.

\bibitem{JSV}R. Brower, P. Rossi and C.-I. Tan, Nucl. Phys. {\bf B190}[FS3]
(1981) 699.\newline A.D. Jackson, M.K. \c{S}ener and J.J.M. Verbaarschot,
Phys. Lett. {\bf B387} (1996) 355.

\bibitem{Jac}J. Verbaarschot, Nato Lectures Cambridge 1997 on ``Confinement, 
Duality, and Nonperturbative Aspects of QCD'' p. 343-378,  hep-th/9710114 

\bibitem{WGW}T. Wilke, T. Guhr and T. Wettig, Phys. Rev. {\bf D57} (1998) 6486.

\bibitem{Jesper}J. Christiansen, Nucl. Phys. {\bf B547} (1999) 329. 

\bibitem{PJZ} P. Zinn-Justin, Commun. Math. Phys. {\bf 194} (1998) 631

\bibitem{ADMN2}G. Akemann, P.H. Damgaard, U. Magnea and S. Nishigaki,
Nucl. Phys. {\bf B519} (1998) 628.

\bibitem{KM3Z}S. Kharchev, A. Marshakov, A. Mironov, A. Morozov and
A. Zabrodin, Nucl. Phys. {\bf B380} (1992) 181.

\bibitem{Dijk}R. Dijkgraaf, hep-th/9201003.

\bibitem{MMS}A. Mironov, A. Morozov and G.W. Semenoff, Int. J. Mod. Phys. 
{\bf A11} (1996) 5031

\bibitem{Zabrodin}A. Zabrodin, solv-int/9704001.

\bibitem{D99}P.H. Damgaard, Nucl. Phys. {\bf B556} (1999) 327.

\bibitem{SmV}A. Smilga and J.J.M. Verbaarschot, Phys. Rev. {\bf D51} (1995) 
829.

\bibitem{D0}P.H. Damgaard, Phys. Lett. {\bf B425} (1998) 151. 

\bibitem{F}T. Nagao and P.J. Forrester, Nucl. Phys. {\bf B435}
(1995) 401.\newline
M.E. Berbenni-Bitsch, A.D. Jackson, S. Meyer, A. Sch\"{a}fer, 
J.J.M. Verbaarschot and T. Wettig, Nucl. Phys. Proc. Suppl. 63 (1998) 820.



\end{thebibliography}
\end{document}